# A Review on MR Based Human Brain Parcellation Methods

Pantea Moghimi[a], Anh The Dang [b], Theoden Netoff[c], Kelvin Lim[d], Gowtham Atluri[b]

[a] Department of Neurobiology, University of Chicago, Chicago, IL
b Department of Electrical Engineering and Computer Science, University of Cincinnati, Cincinnati, OH
[c] Department of Biomedical Engineering, University of Minnesota, Minneapolis, MN
[d] Department of Psychiatry, University of Minnesota, Minneapolis, MN

## Abstract

Brain parcellations play a ubiquitous role in the analysis of magnetic resonance imaging (MRI) datasets. Over 100 years of research has been conducted in pursuit of an ideal brain parcellation. Different methods have been developed and studied for constructing brain parcellations using different imaging modalities. More recently, several data-driven parcellation methods have been adopted from data mining, machine learning, and statistics communities. With contributions from different scientific fields, there is a rich body of literature that needs to be examined to appreciate the breadth of existing research and the gaps that need to be investigated. In this work, we review the large body of in vivo brain parcellation research spanning different neuroimaging modalities and methods. A key contribution of this work is a semantic organization of this large body of work into different taxonomies, making it easy to understand the breadth and depth of the brain parcellation literature. Specifically, we categorized the existing parcellations into three groups: Anatomical parcellations, functional parcellations, and structural parcellations which are constructed using T1-weighted MRI, functional MRI (fMRI), and diffusion-weighted imaging (DWI) datasets, respectively. We provide a multi-level taxonomy of different methods studied in each of these categories, compare their relative strengths and weaknesses, and highlight the challenges currently faced for the development of brain parcellations.

## 1. Introduction

Brain parcellation involves dividing gray-matter locations in the brain into parcels, sets of distinct ideally spatially contiguous locations, such that locations within each parcel (also referred to as a region) share similarity in one or more properties including cyto- or myelo-architecture, their pattern of connectivity to other locations, the functional activity they exhibit, and their topographical representation (Felleman and Van Essen 1991). Brain parcellation maps (also referred to as parcellations) are widely used in studies involving different in vivo imaging modalities. For example, brain parcellations are used in numerous magnetic resonance imaging (MRI) studies that measure the thickness and volume of various brain regions from structural T1-weighted images (e.g. (Aron et al. 2003; Tae et al. 2008)). Brain parcellations are also used by functional MRI (fMRI) studies to report the locus of activity in a given task (e.g. (Nieto-Castanon et al. 2003; Poldrack 2007)). Another common application of brain parcellations is for the construction of network models of the brain using fMRI and diffusion-weighted imaging (DWI) datasets (e.g. ). Brain parcellations have also been used in studies that jointly analyze data from imaging modalities such as magnetoencephalogram (MEG) (Hillebrand et al. 2012) and electroencephalogram (EEG) (Nguyen and Cunnington 2014) in combination with MRI imaging for localizing activity observed on surface electrodes.

Determining a gold standard for brain parcellation is a problem that has captured the interest of neuroscientists for over a century. At the start of the twentieth century, several histological studies, the most prominent of which was that of Korbinian Brodmann, parcellated the human cerebral cortex based on cytoarchitectonics . Resulting regions shared a similar microstructure, i.e., distribution and arrangement of cell bodies in the grey matter and structure of cortical layers. Several histological studies over the span of the next four decades produced numerous parcellations of the cerebral cortex (von Economo and Koskinas 1925; Sarkisov, Filimonoff, and Preobrashenskaya 1949; Bailey and von Bonin 1951). The arrival of the magnetic resonance imaging technology enabled researchers to study the human brain in vivo. Since the microstructure of the cortex is not visible in T1-weighted MRI images, parcellation methods for MR images rely on macrostructures of the brain such as certain sulci and gyri to manually delineate each region (Geyer et al. 2011). Such an approach is expected to result in brain regions with voxels that share similar cytoarchitecture (Talairach and Tournoux 1988; Caviness et al. 1996). Since this procedure is very time consuming, several attempts were made to automate the parcellation procedures using computational algorithms (e.g. (Lancaster et al. 2000; Tzourio-Mazoyer et al. 2002)). With recent advancements in fMRI technology, there have been several attempts to parcellate the brain into regions that are homogenous in functional activity. Using time-series obtained from blood-oxygen-level-dependent (BOLD) signal change, voxels that share similar activation profiles have been grouped using data-driven methods (e.g. (Cohen et al. 2008; Craddock et al. 2012; Blumensath et al. 2013)). Later on, with the availability of DWI data, parcellations were constructed with the goal of grouping locations with similar white matter trajectories (e.g. (Moreno-Dominguez, Anwander, and Knösche 2014; Parisot et al. 2015; Gallardo et al. 2017)).

In the past two decades, several reliable brain parcellations have been developed. These parcellations have become an indispensable tool for analyzing in vivo imaging datasets. Different parcellation methods have been developed by a diverse group of experts including neuroanatomists, neuroscientists, engineers, and data mining experts. With contributions from different communities within and outside neuroscience, there is a rich body of literature published in different venues that needs to be examined to appreciate the breadth of existing parcellations, their relative strengths and weaknesses, and gaps that need to be investigated. In this work, we review existing literature spanning different MR imaging modalities and methods. A key contribution of this work is the organization of this large body of work into different taxonomies, making it easy to provide an overview of the literature. Specifically, we categorized the available brain parcellations into three groups: Anatomical parcellations, functional parcellations, and structural parcellations which are constructed using T1-weighted MRI, fMRI, and DWI datasets, respectively (Figure 1). We provided a multi-level taxonomy of different methods studied in each of these categories. Our main focus is on the methodological aspects of brain parcellations. Our goal is to guide the readers in choosing a brain parcellation method that suits their specific research needs. For broader reviews of brain parcellations that are not limited to MRI based methods the interested reader is referred to (Eickhoff, Yeo, and Genon 2018; Eickhoff, Constable, and Yeo 2018).

Terms 'parcellation', 'atlas', and 'template' have been used in the literature interchangeably (e.g. (Van Essen 2005; Wu et al. 2007)). For the sake of clarification, we refer to any partitioning of the gray matter that specifies what region each gray matter voxel belongs to as a brain 'parcellation'. In the case of anatomical parcellations, each region is assigned a label (e.g. superior temporal gyrus) by a neuroanatomist. We refer to such a labeled brain parcellation as a brain 'atlas'. 'Template' or 'stereotaxic space' on the other hand we refer to a representative T1-weighted MRI brain image used as a standard reference image for registration of other T1-weighted images (e.g. (Holmes et al. 1998)). We also note that while spatial contiguity of regions of a parcellation is a desired property, in this manuscript the term

'region' does not necessarily refer to contiguous groups of voxels, but can also refer to fragmented partitions of the gray matter, which are also referred to as brain 'networks' (e.g. (Power et al. 2011)).

The scope of this article will be parcellation methods developed for the segmentation of adult healthy human brains, with a specific focus on the whole cerebral cortex, and not the parcellation of white matter (e.g. (Makris et al. 1999; Oishi et al. 2008)), neonatal brain (e.g. (Nishida et al. 2006)) and pediatric brain (e.g. (Kaplan et al. 1997; Kates et al. 1999; Shan et al. 2006)). When relevant, we will discuss parcellations of specific parts of the brain, referred to as 'partial parcellations'. An example of a partial brain parcellation is a partitioning of the insular cortex into several regions (Chang et al. 2013). Partial brain parcellations are not the primary focus of this manuscript, however.

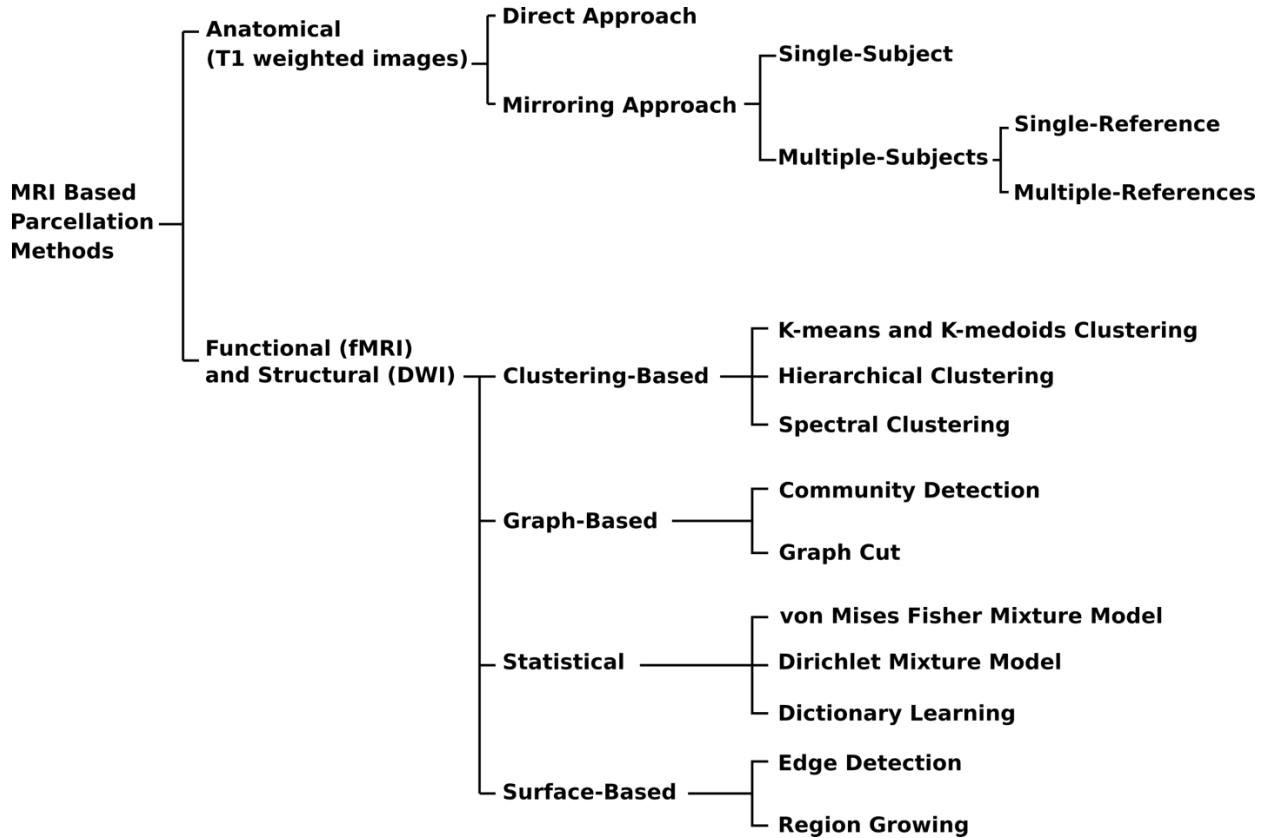

**Figure 1. Taxonomy of MRI based parcellation methods.** We have categorized brain parcellations into two major categories based on the imaging modality used for their construction. Each category is further subdivided into subcategories based on the specific approaches and assumptions of different parcellation algorithms.

## 2. Anatomical parcellations

We define anatomical parcellations as parcellations that are obtained by partitioning the brain into contiguous regions using T1 weighted MR images. A T1 weighted MR image visualizes the folding pattern of the cortical surface, i.e. its sulci and gyri as well as subcortical structures (Hashemi, Bradley, and Lisanti 2010). The sulci and gyri are used by neuroanatomy experts to parcellate the cortex. Each region is then assigned a label that reflects its spatial location. Such a labeled parcellation is also referred to as a brain atlas. The resultant anatomical parcellations are predominantly used as a reference system to report the locus of task-evoked activity in fMRI or PET studies (e.g. (Poldrack 2007; Bartra, McGuire, and Kable 2013; Schecklmann et al. 2013)) and position of brain lesions (e.g. (Schwartz et al. 2009; Wan et al. 2014)) in terms of anatomical regions as opposed to (x,y,z) coordinates that are difficult to interpret. In addition, anatomical parcellations are used for measuring the volume of different brain regions for studies that investigate the morphology of the brain. Examples include studies investigating the effect of aging on brain morphology (e.g. (Allen et al. 2005)) and characterizing changes in brain morphology due to neurological or mental illnesses (e.g. (Wible et al. 1997)).

Anatomical parcellations typically partition the cortex into its major gyri or subdivisions of them. The fundus of major cortical sulci surrounding each gyrus are used as the boundaries of regions. Central to the construction of an anatomical parcellation are the parcellation guidelines, which are developed by neuroanatomy experts. Guidelines consist of four components: 1) a list of necessary macrostructural landmarks, 2) how the landmarks are identified, 3) definition of regions with respect to the listed landmarks, and 4) a set of labels to identify each region. Landmarks mostly comprise major sulci but also include other structures such as poles of the cerebral cortex and subcortical structures such as the Corpus Callosum.

Given a specific set of guidelines, there are various ways to parcellate new brain images. We have categorized different approaches to anatomical brain parcellation into two main categories: direct approach, and mirroring approach (Figure 1). In the following, we will discuss each approach and their relative strengths and weaknesses. A summary of available anatomical parcellations is provided in Table 1.

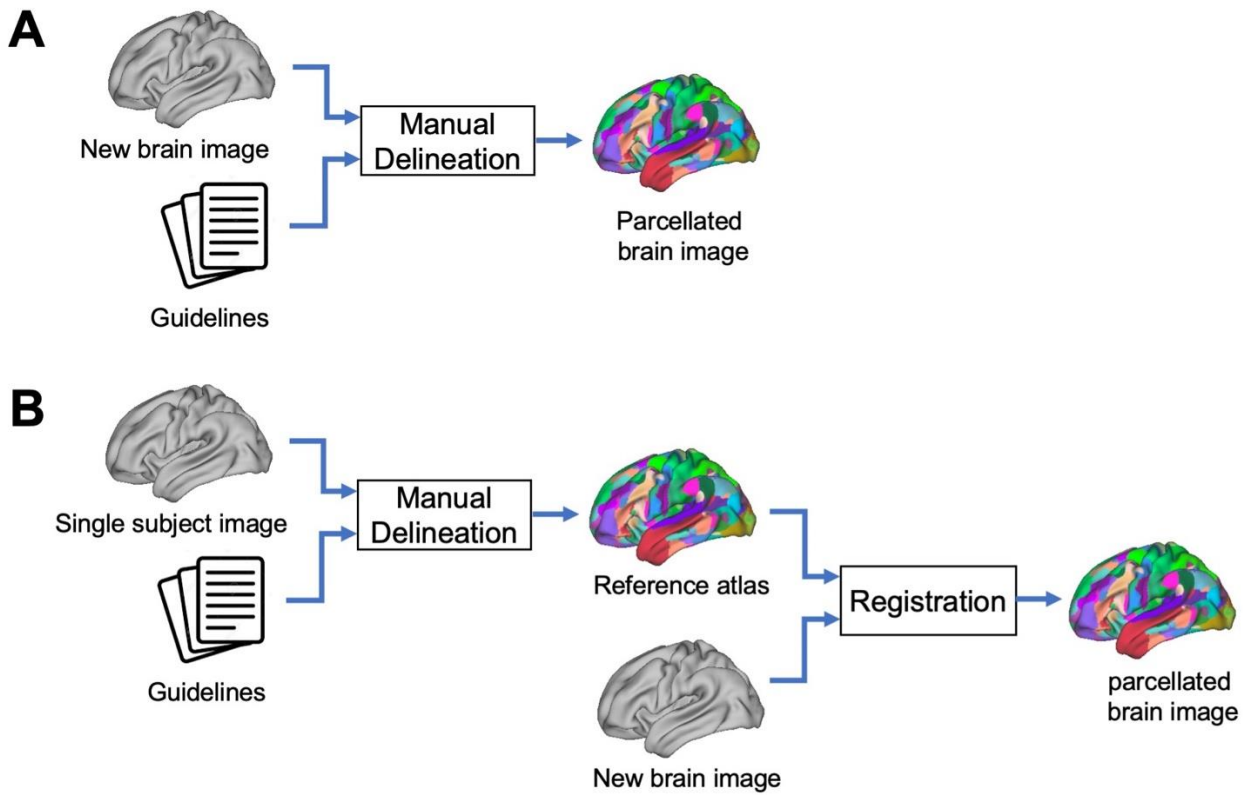

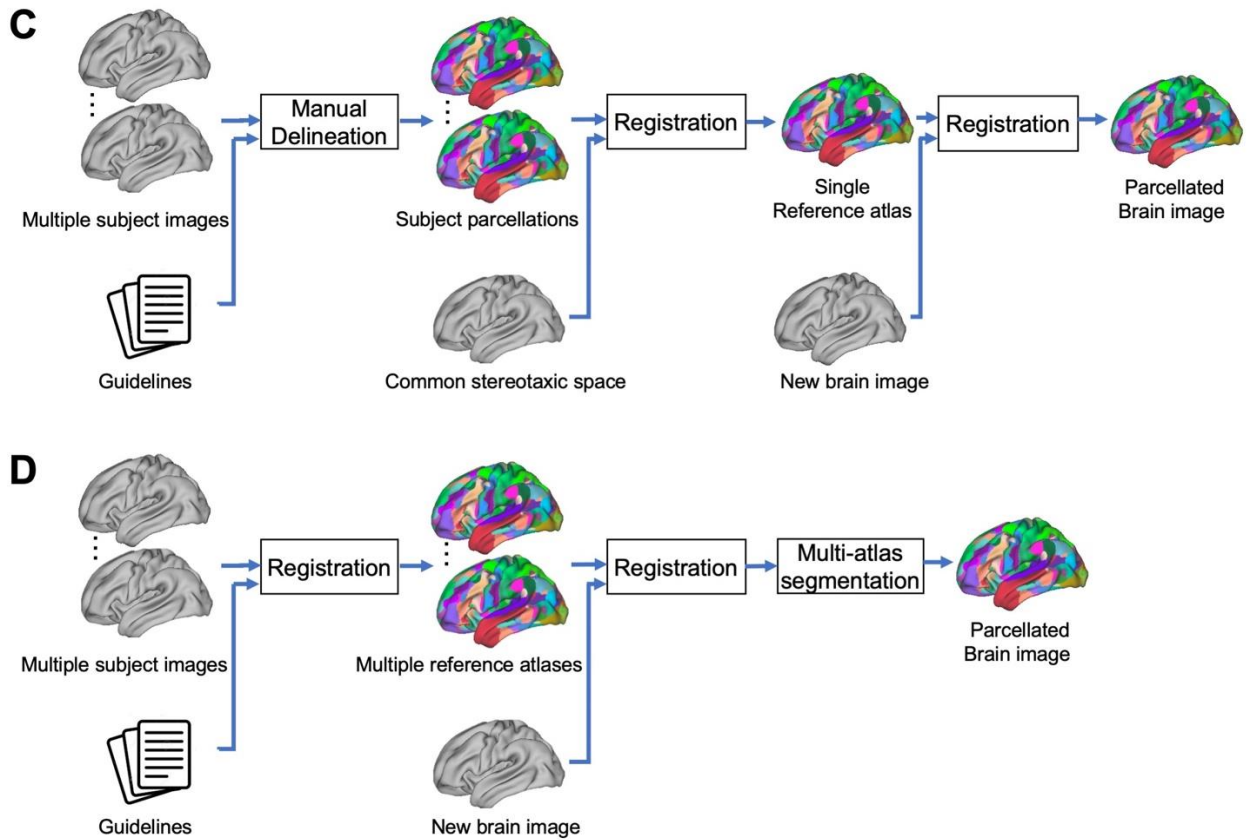

**Figure 2. Schematic of different approaches to the construction of anatomical parcellations.** A) Direct approach; B) Mirroring approach, single-subject; C) Mirroring approach, multiple-subjects, single-reference atlas; D) Mirroring approach, multiple-subjects, multiple-reference atlases.

## 2.1 Direct Approach

The direct approach refers to the manual application of a set of guidelines by an expert to a T1 weighted MR image. This process requires the expert to identify major sulci and their starting and ending points as specified by the guidelines, as well as any other necessary structures. Regions are then delineated based on their bordering landmarks (Figure 2A). The first generation of anatomical parcellations was constructed using the direct approach (Rademacher et al. 1992; Caviness et al. 1996; Tzourio et al. 1997; Crespo-Facorro et al. 2000; Hammers et al. 2002). The direct approach requires the expert to manually parcellate each image, which is labor-intensive and time-consuming (Crespo-Facorro et al. 2000; Shattuck et al. 2008). More recent fMRI data collection efforts in the form of Human Connectome Project (Glasser et al. 2013) and UK Biobank (Miller et al. 2016) that collect thousands of scans render this direct approach obsolete.

## 2.2 Mirroring Approach

The mirroring approach, an alternative to the direct approach, registers a new brain image that is to be parcellated to a brain image that has been manually parcellated, a process also known as 'spatial normalization'. After registration, the label assigned to each voxel from the parcellated image is transferred to its corresponding voxel in the new image. Since the assignment of voxels in the new registered image mirrors that of the existing parcellation, we refer to this approach as the mirroring approach and the manually constructed parcellation as the 'reference atlas'. The mirroring approach provides the user with the reference atlas, and new brain images are parcellated by only performing image registration, for which mature, automated, and efficient techniques exist (Oliveira and Tavares 2014). The mirroring approach eliminates the need for manual labor and hence has become the standard practice for constructing anatomical parcellations.

Reference atlases are constructed by manually parcellating T1-weighted images of one or more subjects. We have categorized reference atlases into 'single-subject' and 'multiple-subjects' based on the number of subjects used for their construction (Figure 1).

**Single-subject-** The first generation of reference atlases was constructed using a single subject and we refer to them as the single-subject atlases (Figure 1). With this approach, a single subject's image is delineated manually by an expert and is treated as the reference atlas. The single subject's image is also referred to as the 'stereotaxic space', or 'template'. Templates are representative brain images that are free of any abnormalities. Commonly used templates are the Talairach template (Talairach and Tournoux 1988), MNI single-subject template (Collins et al. 1998), and the MNI152 template (Fonov et al. 2011). A new brain image that needs to be parcellated is first registered to the reference atlas. Each voxel then adopts the label of the corresponding voxel in the reference atlas (Figure 2B). Popular single-reference atlases include the Talairach atlas (Talairach and Tournoux 1988; Lancaster et al. 1997, 2000), the 'automated anatomical labeling' atlas (AAL) (Tzourio-Mazoyer et al. 2002), the AAL2 atlas (Rolls, Joliot, and Tzourio-Mazoyer 2015), the ICBM atlas , and the temporal atlas developed by Hammers and colleagues (Hammers et al. 2002). Reference atlases constructed using a single subject's brain, however, were found to be biased to unique morphology and idiosyncrasies of that brain (Hammers et al. 2003; Fischl et al. 2004), prompting the field to develop reference atlases using multiple subjects.

**Multiple-subjects-** The multiple-subjects approach uses manually constructed parcellations of T1-weighted images from multiple subjects for partitioning new brain images. The use of multiple subjects reduces the biases introduced by each subject's unique brain anatomy. The label assigned to each voxel across the manually parcellated brains is used to parcellate the new brain image in a process known as multi-atlas segmentation (Iglesias and Sabuncu 2015).

Anatomical parcellations obtained using the multiple-subjects approach are more accurate compared to parcellations obtained using the single-subject approach, where accuracy is defined as how well the resultant anatomical parcellation of an image agrees with a manual parcellation of it (Heckemann et al. 2006; Iglesias and Sabuncu 2015). The multiple-subjects approach has the additional benefit of providing a probabilistic representation of the anatomy of the brain, where statistical models are used to estimate the probability of a given voxel belonging to a region based on the voxel's assignment across multiple subjects. These models quantify how confident the parcellation procedure is in each voxel's assignment. The multiple-subjects approach also allows for quantification of inter-subject variability in morphology of the brain and its spatial pattern, i.e. morphology of which parts of the brain is more variable across subjects and to what extent .

Two approaches are being used to parcellate a new brain image using multiple manual parcellations. One approach registers all the subject parcellations to a common stereotaxic space and constructs a single reference atlas (Figure 2C). We refer to this approach as the 'single-reference atlas' (Figure 1). Each voxel within the reference atlas is assigned a probability density over brain regions. In other words, the reference atlas specifies the probability of each voxel belonging to each brain region. One method to calculate the probability density is to calculate the normalized frequency of the voxel assignments across all manual parcellations. A new brain image that needs to be parcellated is registered to the reference atlas. Each voxel in the new brain image is then assigned to the region with the highest probability at that voxel in the reference atlas. This method is known as majority voting since each of the subject parcellations used for the construction of the reference atlas casts a vote regarding a given voxel's label and the region with the highest number of votes wins. Another method to calculate the probability density is to use statistical models to calculate the probability of each voxel belonging to each region using the probability of it belonging to that region in the reference atlas in addition to i) surface geometry of the brain at that voxel for parcellation of the cortex (Fischl et al. 2004), or ii) image intensity at the neighborhood of that voxel for parcellation of subcortical structures (Fischl et al. 2002). This method is currently implemented in FreeSurfer and the Human Connectome Project image processing pipeline (Glasser et al. 2013). The weakness of the single-reference atlas approach is that it requires a single image registration between the reference atlas and the new brain image that is to be parcellated, and image registration is prone to errors. Inaccuracies in the image registration process propagate to the parcellation result and reduce parcellation accuracy (Klein and Hirsch 2005; Heckemann et al. 2006; Shattuck et al. 2008; Allen et al. 2008; Yeo et al. 2008; Klein et al. 2009).

The second approach which we refer to as 'multiple-reference atlases' (Figure 2D) addresses the above problem by registering each manually constructed parcellation to the new brain image separately instead of constructing a single reference atlas . Each voxel of the new brain image is assigned to a region based on the assignment of its corresponding voxel at each of the multiple reference atlases . Voxel assignments across the reference atlases are fused using multi-atlas segmentation methods to parcellate a new brain image (Iglesias and Sabuncu 2015).

The simplest and probably most popular method of atlas fusion is majority voting, where a voxel is assigned to the region it is most frequently assigned to across all reference atlases (Klein et al. 2005; Heckemann et al. 2006). However, majority voting weighs different reference atlases equally. Different reference atlases are not equally similar in their morphology to the new brain image that is to be parcellated. For example, the new brain image might be missing a sulcus that is present in some reference atlases and absent in others. As a result, it makes sense to only select and use reference atlases that are more similar to the brain image one wishes to parcellate. In fact, it has been shown that using only the reference atlases that are morphologically similar to the new brain image enhances parcellation accuracy (Heckemann et al. 2006; Aljabar et al. 2009). Comparable improvement in parcellation accuracy has also been reported when only reference atlases obtained from subjects whose age was close to the age of the subject whose new brain image was used for parcellation (Aljabar et al. 2009), although combining similarity in both age and morphology to select reference atlases results in further improvements (Lötjönen et al. 2010). Selection based on age is the less computationally expensive of the two methods because matching morphology of the reference atlas requires registering the new brain image to all available reference atlases and choosing the atlases with higher image registration quality. Irrespective of the selection method, reference atlas selection is an effective way to deal with inter-subject differences in brain morphology and increase parcellation quality. To this end, several databases of reference atlases have been released recently. Given a new brain image, the database chooses a set of suitable reference atlases based on the new subject's age or brain morphology (Heckemann et al. 2015). The size of the database has to be large enough to encompass a wide range of brain morphologies and subject idiosyncrasies. Larger variability across reference atlases of a given database has been shown to improve parcellation accuracy since it increases the probability of finding a close match to the new brain image among the available atlases .

When selecting reference atlases based on age, atlases are divided into several age groups. Younger age groups are typically narrower (spanning 4 years) than older age groups (spanning 10 -20 years) due to the developmental changes in the brain morphology of younger subjects. A new brain image is registered only to reference atlases that are within the same age group as the new subject.

Atlas selection based on morphology can be done locally or globally. Global atlas selection algorithms choose a set of reference atlases that are the most similar to the new brain image in their overall morphology (Lötjönen et al. 2010; Heckemann et al. 2015). The morphological similarity is measured as the performance of image registration between the new brain image and each of the reference atlases. Local atlas selection methods, on the other hand, choose a set of atlases for each brain region separately. Local registration performance within the vicinity of each region is used to select reference atlases. The rationale is that while a reference atlas might be an overall good match for a given brain image, it might not be the optimal match for every brain region, for example, because the atlas is missing a sulcus that is present in the vicinity of the region one wishes to delineate. Although being more computationally expensive, this method is more suitable for studies that aim at delineating a single region only, for example, epilepsy studies that need to delineate the hippocampus.

The selection of only the most similar reference atlases to the new brain image is the equivalent of assigning binary weights to each reference atlas, where selected atlases are assigned a weight of one and excluded atlases zero. An alternative approach is to fuse reference atlases using non-binary weights, where reference atlases that are more similar to the new brain image are assigned higher weights. The advantage of this approach is that all reference atlases will be used, maximizing the amount of information that is used for parcellation. How different atlases are weighted is the subject of active research, details of which are beyond the scope of this manuscript. The interested reader is referred to (Iglesias and Sabuncu 2015). Similar to atlas selection, weight assignment can also be done globally (Warfield, Zou, and Wells 2004; Artaechevarria, Munoz-Barrutia, and Ortiz-de-Solorzano 2009; Sdika 2010; Asman and Landman 2013) or locally. Global weights are assigned based on the global similarity between each reference atlas and the new brain image and are equal across all voxels. Local weights are assigned based on the similarity between the reference atlas and the new brain image in a local neighborhood of each voxel and differ from voxel to voxel. Several studies have shown that weights assigned based on local similarity result in more accurate parcellations than globally assigned weights (Artaechevarria, Munoz-Barrutia, and Ortiz-de-Solorzano 2009; Sabuncu et al. 2010), although this improvement comes at the cost of computation time.

An important question is how many reference atlases are required to accurately parcellate a new brain image. The relationship between parcellation accuracy and the number of atlases is not monotonic (Aljabar et al. 2009; Lötjönen et al. 2010). Parcellation accuracy increases as more reference atlases are used for parcellation and reaches its peak after the inclusion of on average 10-20 atlases (Aljabar et al. 2009). The inclusion of more atlases results in a small decline in accuracy because atlases are added based on their rank order in terms of their similarity to the new brain image. The inclusion of more atlases entails the inclusion of atlases that are less similar to the new brain image which results in a reduction in parcellation accuracy.

Recent advances in multi-atlas segmentation methods have proven promising and their use has substantially improved parcellation quality. Use of software packages that have implemented these methods and provide a database of reference atlases are preferred.

## 2.3 Additional remarks

The primary challenge for anatomical parcellation of the brain is caused by inter-subject variability in the morphology of the cortical surface. Variations in the position and direction of sulci make identification of some sulci difficult, and some sulci are not found in every subject (Ono, Kubik, and Abernathey 1990). Several guidelines have dealt with this issue by defining arbitrary conventions to keep manual parcellation consistent across experts (e.g.(Crespo-Facorro et al. 2000; Hammers et al. 2003)). For example identification guidelines for paracentral sulcus as provided by Crespo-Facorro and colleagues say: "We arbitrarily defined the paracentral sulcus as the nearest sulcus posterior to an orthogonal line that passes through the anterior commissure and is perpendicular to the anterior commissure posterior commissure (AC-PC) line" (Crespo-Facorro et al. 2000). Another approach is to group regions separated by problematic sulci together, obviating the need to identify the sulcus altogether. For example, Destrieux and colleagues grouped the 4 orbital gyri into a single region, because the orbital sulcus that separated them could not be consistently identified across subjects (Destrieux et al. 2010). Either approach relies on the neuroanatomist's judgment of how to delineate regions rather than a gold standard that is globally accepted upon. The subjective bias introduced in the process, however, can be considerable. For example, the volume of the hippocampus has been reported to range from 2.4-5.3 cm$^3$ depending on how conservative the guidelines are about the borders of the hippocampus (Geuze, Vermetten, and Bremner 2005; Boccardi et al. 2011). Such variability has serious consequences when it comes to interpreting and comparing results of studies that use different anatomical parcellations. For example, the extent of atrophy in hippocampal volume caused by Alzheimer's disease might be different across studies merely because of how the hippocampal region is delineated, making it difficult to directly compare studies that use different anatomical parcellation protocols. With recent advancements in the MRI technology and the availability of high-resolution images that make the tracing of sulci easier, such conventions can potentially be revisited and improved upon.

Another challenge in the identification of the landmarks is the subjective nature of manual parcellation. Different experts have different judgments about the position of landmarks for parcellation and might make mistakes in identification of the boundaries, leading to discrepancies between parcellations of the same brain, created by different experts (Caviness et al. 1996; Heckemann et al. 2006). To reduce the impact of such mistakes, automated algorithms that use different image processing techniques to identify the sulci can be used (e.g. ). In fact, a few anatomical parcellation methods exist that use these sulci identification techniques to automatically delineate major gyri . Automatization of sulcus detection eliminates potential subjective biases of the neuroanatomist and errors introduced by the registration process (Auzias et al. 2016). However, to the best of our knowledge, these techniques have not yet been implemented in the common MRI software packages.

| Atlas Name | Brief Description |
|---|---|
| Harvard-Oxford (HO) | The set of guidelines for the construction of this atlas were developed in 1992 by Rademacher and colleagues and were later used to construct a **multiple-subjects, single-reference atlas** in the MNI152 stereotaxic space (Rademacher et al. 1992; Caviness et al. 1996; Fischl et al. 2004). <br> **Number of Regions:** 117, includes the cortex, basal ganglia, subcortical structures, and ventricles <br> **Subject(s):** 37 healthy subjects, ages 18-50 years old |
| Crespo-Facorro and colleagues | The guidelines for this atlas were developed for **direct parcellation** of the brain with the aim of improving parcellation precision by providing details on how to trace major sulci across consecutive slices (Crespo-Facorro et al. 2000). <br> **Number of Regions:** 82, includes only the cortex |

| Li and colleagues | The set of guidelines by Crespo-Facorro and colleagues in 2000 were refined to merge smaller regions that were difficult to delineate in all subjects. The guidelines were used to produce a **multiple-subjects, single-reference atlas**. Brain image from one of the subjects was used as the stereotaxic space (W. Li et al. 2013).<br>**Number of Regions:** 48, includes only the cortex<br>**Subject(s):** 25 healthy subjects, 25 subjects with schizophrenia, 12-41 years old |
|---|---|
| Talairach Daemon | This atlas is based on the original Talairach atlas and includes 5 parcellation levels: hemispheric, lobular, gyrus, tissue, and cell level. The gyrus level parcellation is the commonly used parcellation that is similar to other anatomical atlases discussed. A **single-subject reference atlas** in the Talairach stereotaxic space is available which is a digitized version of the manual parcellation done by Talairach and Tournoux in 1988 (Talairach and Tournoux 1988; Lancaster et al. 2000).<br>**Number of Regions:** 110, includes the cortex, subcortical structures, cerebellum, and ventricles<br>**Subject(s):** Postmortem MRI images of 1 alcoholic subject, 60 years old, known as the 'Talairach template' |
| AAL | The guidelines for this atlas were developed in 1997 by Tzourio and colleagues and later applied to parcellate the MNI single-subject stereotaxic space to produce a **single-subject reference atlas** (Tzourio et al. 1997; Tzourio-Mazoyer et al. 2002).<br>**Number of Regions:** 116, includes the cortex, basal ganglia, subcortical structures, and cerebellum<br>**Subject(s):** 1 healthy subject |
| AAL2 | This atlas is a revision of the AAL atlas, where the definition of regions located in the orbitofrontal cortex was modified to facilitate comparison of the orbitofrontal cortex anatomy between humans and macaque monkeys. Updated guidelines were applied to parcellate the MNI single-subject stereotaxic space to produce a **single-subject reference atlas** (Rolls, Joliot, and Tzourio-Mazoyer 2015).<br>**Number of Regions:** 120, includes the cortex, basal ganglia, subcortical structures, and cerebellum<br>**Subject(s):** 1 healthy subject |
| ICBM single-subject atlas | This atlas includes subdivisions of the thalamus and basal ganglia. The guidelines were applied to parcellate the MNI single-subject stereotaxic space to produce a **single-subject reference atlas** (Kabani, Collins, and Evans 1998; J. Mazziotta et al. 2001).<br>**Number of Regions:** 122, includes the cortex, basal ganglia, subcortical structures, brainstem, and cerebellum<br>**Subject(s):** 1 healthy subject |
| Hammers, single-subject | The guidelines for this atlas were developed with a focus on the temporal lobe, making it suitable for studies that acquire images in temporal lobe orientation such as epilepsy. They were used to parcellate the MNI single-subject stereotaxic space to produce a **single-subject reference atlas** (Hammers et al. 2002).<br>**Number of Regions:** 41, includes the cortex, basal ganglia, subcortical structures, and cerebellum<br>**Subject(s):** 1 healthy subject |
| Hammers, multiple-subjects | The guidelines previously developed by Hammer and colleagues (Hammers et al. 2002) were expanded to include brain stem and ventricles. They were used to construct a **multiple-subjects, single-reference atlas** in the MNI152 stereotaxic space (Hammers et al. 2003).<br>**Number of Regions:** 49, includes the cortex, basal ganglia, subcortical structures, cerebellum, brain stem, and ventricles<br>**Subject(s):** 20 healthy subjects, median age 30.5 |
| Heckemann and colleagues | The guidelines developed by Hammers and colleagues (Hammers et al. 2003) were augmented to produce a finer parcellation of the cortex. The guidelines were used to construct **multiple-subjects, multiple-reference atlases** (Heckemann et al. 2006, 2015).<br>**Number of Regions:** 67, includes the cortex, basal ganglia, subcortical structures, ventricles, and cerebellum<br>**Subject(s):** 30 healthy subjects, 20-54 years old |
| LPBA40 | The guidelines for this atlas were used to construct three **multiple-subjects, single-reference atlases** in the ICBM452 1mm resolution, ICBM452 2mm resolution, and ICBM152 stereotaxic spaces (Shattuck et al. 2008).<br>**Number of Regions:** 56, includes the cortex, basal ganglia, cerebellum, and brain stem<br>**Subject(s):** 40 healthy subjects, 19-40 years old |
| Desikan-Killiany (DK) | This **multiple-subjects, single-reference atlas** was constructed using a group of subjects of a wide age range that also included patients with brain atrophy due to Alzheimer's disease. Such a selection of subjects was to make this atlas applicable to different groups of people of different ages or at different brain atrophy stages. This atlas is available in the spherical coordinates used by the FreeSurfer software package (Desikan et al. 2006).<br>**Number of Regions:** 68, includes only the cortex<br>**Subject(s):** 30 healthy subjects, 10 patients with Alzheimer's disease, 19-87 years old |
| Desikan-Killianny-Tourville (DKT) | The guidelines from the DK atlas were revised and some regions whose boundaries could not be identified reliably were incorporated into other regions. A **multiple-subjects, single-reference atlas** was constructed. A variant of this atlas with 50 regions is also available, where regions that were subdivisions of the major gyri were combined (Klein and Tourville 2012).<br>**Number of Regions:** 62, includes only the cortex<br>**Subject(s):** 40 healthy subjects, average 26 years old |
| Destrieux | This **multiple-subjects, single-reference atlas** divided the cortex into regions that were either a gyrus or a sulcus, as opposed to other anatomical atlases, where all cortical regions were gyri, whose boundaries were at the fundus of their bordering sulci. Voxels visible on the pial surface of the cortex were assigned to gyral regions and voxels on the banks of the cortical sulci were assigned to sulcal regions (Destrieux et al. 2010).<br>**Number of Regions:** 74, includes only the cortex<br>**Subject(s):** 24 healthy subjects, 18-33 years old |

**Table 1. Summary of available anatomical atlases.** The number of regions for each atlas includes both hemispheres. The age of subjects used for the construction of reference atlases is reported when the information was available.

Different reference atlases parcellate the brain using different guidelines and region definitions. Comparing the results of studies that use different atlases requires establishing concordance between the used atlases. Working out concordance between regions of different atlases is not straightforward, however. Some regions with the same label have different definitions across different atlases. For example, the definition of caudate in the LPBA40 atlas includes voxels that belong to both caudate and accumbens in the Harvard-Oxford atlas (Bohland et al. 2009). Establishing concordance between two different anatomical atlases becomes even more complex when a single region from one atlas is further divided into multiple sub-regions in another atlas, or voxels belonging to a set of regions in one atlas correspond to voxels belonging to another set of regions in a different atlas. These relationships have been characterized for several (but not all) commonly used anatomical atlases and can be used for comparing results from studies that have used different anatomical atlases (Bohland et al. 2009). More work is however required to characterize atlases not included by Bohland and colleagues.

## 3. Functional and structural parcellations

Functional and structural parcellations are partitionings of the brain constructed using functional magnetic resonance imaging (fMRI) and diffusion-weighted imaging (DWI) datasets respectively. FMRI data captures the functional activity of voxels quantified as local changes in the oxygen level of neuronal tissue, also referred to as the BOLD (Blood-oxygen-level-dependent) signal (Figure 3A) (Ogawa et al. 1990, 1992). DWI data estimates the location and orientation of white matter fibers in the brain. It measures the direction of diffusion of water molecules in the neural tissue (Figure 3D). The fact that the diffusion of water molecules is higher along the direction of white matter tracts is used to estimate the orientation of fibers at any given voxel and reconstruct the major white matter fiber bundles (Alexander et al. 2007; Soares et al. 2013).

Functional parcellations are constructed by grouping voxels with similar functional properties. The two functional properties used for parcellation are functional activity (Figure 3B) and functional connectivity profile of voxels (Figure 3C). Functional activity is the fluctuations in the BOLD signal over time and recorded as a single time-series per voxel. The functional connectivity profile of a voxel, also known as its functional connectivity fingerprint, is the degree of joint variability of the voxels' time-series with every other voxel in the brain, typically quantified using the Pearson correlation coefficient (Friston 2011). Functionally distinct brain regions are assumed to consist of voxels with shared functional activity and functional connectivity profiles (Passingham, Stephan, and Kötter 2002). By grouping voxels that share similar functional properties together, the functional parcellation is expected to result in functionally segregated regions.

Two types of fMRI data are typically used for functional parcellation: Resting-state fMRI and Task fMRI. Resting-state fMRI, also known as intrinsic functional activity, is collected while the subject is resting in the scanner, whereas task fMRI is collected while the subject is engaged in tasks such as moving a hand or foot, watching visual stimuli, or more cognitively demanding tasks such as working memory tasks. The use of resting-state fMRI for functional parcellation is more common since it captures spontaneous fluctuations in neural activity and is hypothesized to reveal fundamental properties of brain organization without selectively engaging any brain regions.

Structural parcellations are constructed by grouping voxels with similar structural connections to other voxels (Figure 3D). A structural connection between a pair of voxels exists if a white matter tract originating from one voxel terminates in the other, i.e. the two voxels are directly connected with a white matter tract. A voxels' structural connectivity profile is the pattern of white matter connections between that voxel and every other voxel in the brain. Voxels with similar structural connectivity profiles are hypothesized to serve the same functional role. The reason is that the functional role of a region is constrained and determined by what inputs and outputs it receives or sends out through its structural connections from or to other regions (Passingham, Stephan, and Kötter 2002). Therefore, parcellating the brain into regions with distinct structural connectivity profiles is hypothesized to produce functionally relevant regions.

Both fMRI and DWI techniques produce large and complex datasets that are not feasible to analyze manually. Grouping of voxels needs to be done using data mining algorithms developed to discover the grouping structure of big datasets. The rich body of data mining literature consists of a plethora of different algorithms. Each algorithm partitions the data into separate groups based on specific assumptions it makes about the dataset's structure. Several of these methods have been used for constructing functional and structural parcellations.

In what follows, we will first discuss how the similarity between voxels is quantified for functional and structural parcellations. We will then briefly describe different parcellation algorithms and compare and contrast the assumptions different parcellation algorithms make about the data.

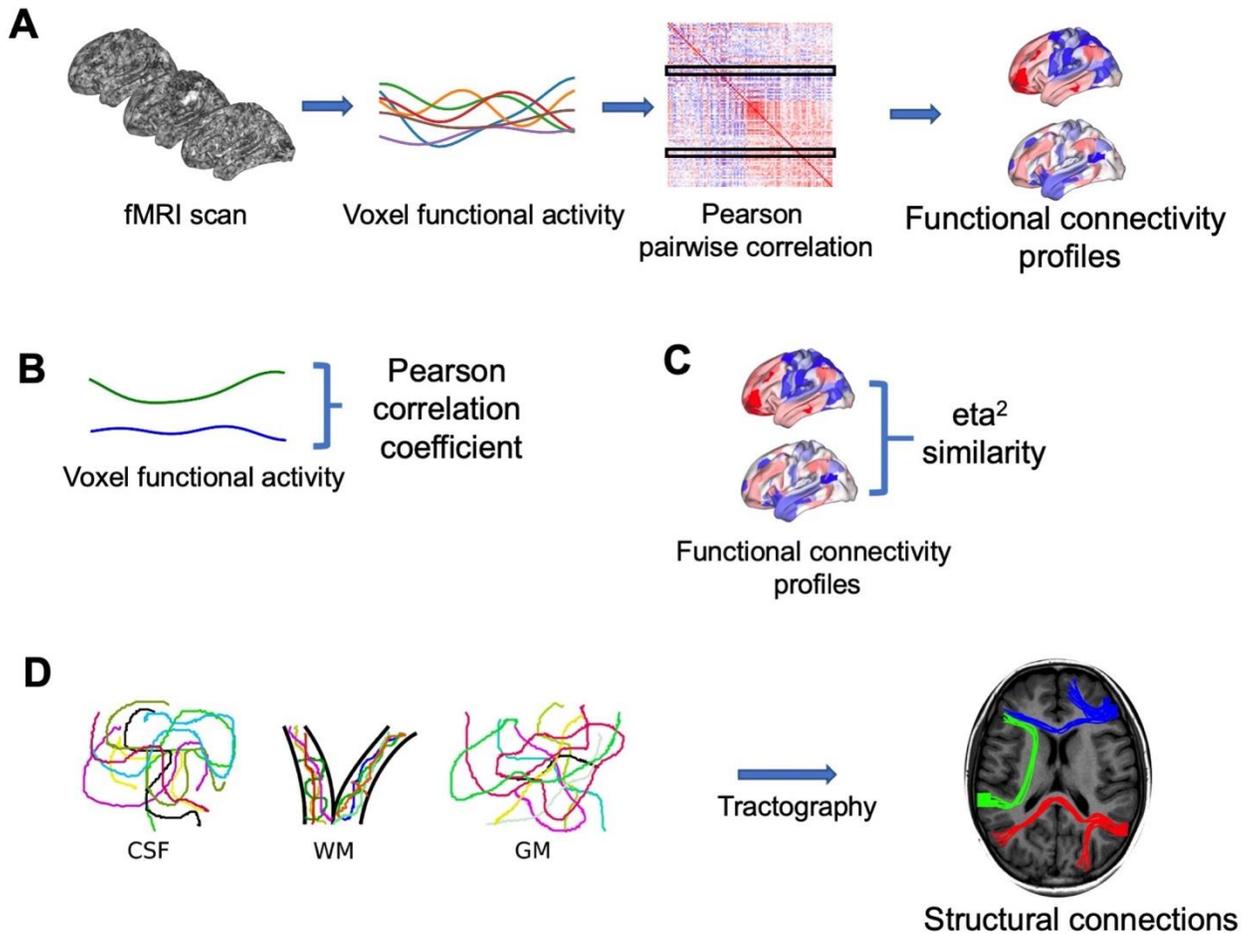

**Figure 3. Similarity between voxels in functional and structural parcellations.** A) FMRI measures brain activity in each voxel captured as fluctuations in the BOLD signal over time. Functional properties of each voxel are estimated from its BOLD signal fluctuations. Voxels with similar functional properties are grouped to form functionally relevant regions. The two functional properties used for construction of functional parcellations are pairwise similarity in functional activity and pairwise similarity in functional connectivity profiles. Functional connectivity profiles are constructed from the Pearson pairwise correlation matrix. Each row of the matrix is the pairwise correlation between a voxel and all other voxels and is treated as the functional connectivity profile of that voxel. Color maps are toy examples of functional connectivity profiles of two voxels and reflect the correlation coefficient between the voxel and every other voxel, B) Pairwise similarity in functional activity between two voxels, measured as the Pearson correlation coefficient between their time-series, C) Pairwise similarity in functional connectivity profiles between a pair of voxel*s* is typically measured using $eta^2$., D) DWI datasets are used to estimate structural connections between voxels using tractography algorithms. DWI measures the Brownian motion of water molecules which is uniformly distributed in all directions in cerebrospinal fluid (CSF) and gray matter (GM) but dominantly along the axonal trajectory in the white matter (WM). Structural parcellations group voxels with similar white matter connectivity profiles to form regions.

## 3.1 Similarity measure

**Functional activity-** The similarity between the functional activity of two voxels (Figure 3B) is quantified as the degree of covariation between their respective time-series measured as Pearson correlation coefficient (Craddock et al. 2012; Blumensath et al. 2013; Shen et al. 2013; Arslan, Parisot, and Rueckert 2015). Euclidean distance has also been used to compute the similarity between time-series (Mezer et al. 2009; Lee et al. 2012). The main difference between Euclidean distance and Pearson correlation coefficient is that Euclidean distance is sensitive to the magnitude of the time-series as well as their joint variation whereas the Pearson correlation coefficient is insensitive to magnitude. The magnitude of the time-series is however affected by irrelevant factors such as the distance of the voxel from major vessels (Penny et al. 2011). Therefore, the use of the Pearson correlation coefficient is preferable and more common to Euclidean distance, unless the time-series are normalized to have zero mean and unit variance, in which case the two measures are equivalent (Jing Wang and Wang 2016). In the case of functional parcellations constructed using task-evoked fMRI, the similarity between voxels has also been quantified using task-relevant measures such as the correlation coefficient between average time-series aligned with respect to task-related events and task-evoked activity across multiple tasks.

**Functional connectivity profile-** The functional connectivity profile of a voxel is the functional connectivity between that voxel and all other voxels in the brain (Figure 3C). Functional connectivity between a pair of voxels is the Pearson correlation coefficient between their time-series (Friston 2011). Functional connectivity profiles can be binarized by thresholding and keeping the strongest connections (Power et al. 2011; Yeo et al. 2011; Najafi et al. 2016). Binarizing is required when the parcellation algorithm is only applicable to binary profiles (e.g. (Power et al. 2011)). Binarization has also been reported to produce better parcellation results by removing spurious weak connections, even when the parcellation algorithm does not require binary profiles (Yeo et al. 2011; Najafi et al. 2016). It should be noted that this enhancement comes at the cost of discarding information and could potentially be specific to the parcellation algorithm. In other words, some parcellation algorithms might be more sensitive to the presence of weak connections than others and would benefit from binarization more considerably. Therefore, the impact of binarization on the quality of parcellation should be tested empirically.

The Pearson correlation coefficient appears to be a promising measure to calculate the similarity between functional connectivity profiles (e.g. ). However, the Pearson correlation coefficient only captures the degree of covariation between the two profiles and does not take into account the magnitude of connectivity in non-binary connectivity profiles. The magnitude of connectivity profiles captures the strength of functional connections between voxels which is relevant information for parcellation. Voxels with different functional roles might be strongly connected to the same areas of the brain and weakly to others but with different connection strengths. Such voxels will have spuriously highly correlated connectivity profiles. Several studies have therefore used the eta$^2$ coefficient to quantify the similarity between connectivity profiles (Cohen et al. 2008; Kelly et al. 2010; Barnes et al. 2010; Kelly et al. 2012; Long et al. 2014; Jung et al. 2014; Gordon, Laumann, Gilmore, et al. 2017). Eta$^2$ coefficient measures the amount of variance in one variable explained by the variance of another variable (Tabachnick and Fidell 2006) and is sensitive to connectivity magnitudes as well as their covariation.

The two functional properties, i.e. similarity between functional activity or functional connectivity profiles, can both be used as similarity measures for functional parcellation. Both measures are hypothesized to have high values for voxels with similar functional roles (Yeo et al. 2011; Craddock et al. 2012) and the parcellation algorithms discussed in this paper (subsection 3.2) can be used in combination with either measure. The two measures are highly correlated but share a nonlinear relationship (Craddock et al. 2012). The question then is which measure to choose for functional parcellation. It has been reported that similarity in functional activity instead of functional connectivity, results in parcellations that have higher consistency across different subjects (Craddock et al. 2012). However, this result should be interpreted with caution and does not indicate that similarity in functional activity produces more accurate parcellation results. The observed results could be method specific. In addition, differences in parcellation results across subjects could be the result of both inter-subject differences and parcellation inaccuracies. Subjects have been shown to have highly distinct functional connectivity profiles (Finn et al. 2015; Gordon, Laumann, Adeyemo, and Petersen 2017) which could explain the observed higher inter-subject variability when similarity in functional connectivity profile was used for parcellation. A more rigorous comparison of the two measures needs to be performed.

**Structural connectivity profile-** The structural connectivity profile of a voxel is calculated using tractography algorithms that estimate the pathway of white matter tracts and where they originate and terminate within the gray matter (Jbabdi and Johansen-Berg 2011). The connectivity profile of a voxel captures which voxels of the brain are directly connected to it via a white matter tract or structural connection. Structural connectivity profiles estimated using deterministic tractography algorithms have a value of one for all voxels that are connected to that voxel via a white matter fiber and zero for all other voxels. If probabilistic tractography is performed to estimate white matter tracts, the connectivity profile of a voxel captures the probability of a fiber connecting the voxel to every other voxel in the brain. The result is a connectivity profile that is not binary but consists of values between zero and one.

The similarity between structural connectivity profiles has been calculated using the Pearson correlation coefficient . As previously discussed, the Pearson correlation coefficient is insensitive to the magnitude of structural connections between voxels. Therefore, it might be a suitable measure when binary tractography is performed. Probabilistic tractography algorithms, on the other hand, produce a non-binary value for each pairwise structural connection which is potentially informative. Hence, eta$^2$ is a better candidate for quantifying similarity between structural connectivity profiles estimated by probabilistic tractography.

To reduce the impact of potential errors in the tractography algorithm, several studies have estimated structural connectivity profiles by calculating structural connectivity between each voxel and a predefined set of 'target regions' instead of other voxels. As a result, connectivity information is aggregated across several voxels within each target region which reduces the impact of any error introduced by the tractography algorithm on the parcellation results (Jin et al. 2015). This method of constructing connectivity profiles has the added benefit of reducing the size of the connectivity profiles and consequently computation time.

Multiple approaches have been used for the definition of target regions: i) for partial parcellation studies, target regions can be regions with previously known structural connections to the part of the brain that is to be parcellated as detected by invasive tracing experiments in non-human primates , ii) target regions can be regions delineated by an available anatomical parcellation , iii) target regions can also be contiguous regions constructed by randomly grouping spatially adjacent voxels (Roca et al. 2009, 2010; Jin et al. 2015). The first approach is not suitable for the construction of whole-brain parcellations since it relies on establishing one to one homology between regions of non-human primate brains and human brains and several histologically delineated regions in the human brain have not been identified in other species (e.g. ). For the construction of whole-brain parcellations, second and third approaches have been used. However, it has been shown that the particular grouping of voxels, i.e. the number of target regions and how they are defined does not significantly affect the resultant structural parcellation (Jin et al. 2015). However, the choice of target regions can be relevant for partial brain parcellations. Jbabdi and colleagues observed that when parcellating a small part of the brain, such as parcellating the thalamus to its subdivisions, certain target regions are more informative than others. For example, when parcellating the thalamus, connectivity profiles of thalamic voxels to sensori-motor and parietal cortices diverged across different modules of the thalamus .

It is worth noting that for parcellation of subcortical nuclei, in addition to structural connectivity profiles, local diffusion properties of voxels can also be used without performing tractography (Wiegell et al. 2003; Solano-Castiella et al. 2010). The reason is that white matter penetration into the gray matter is significantly deeper in subcortical structures compared to the cortex. Over the cortical surface diffusion properties are relatively homogenous and can't be used for parcellation (Behrens, Woolrich, et al. 2003).

## 3.2 Parcellation algorithm

A variety of algorithms have been used to construct functional and structural parcellations. We have categorized these algorithms into 4 categories based on the representation of data used for each approach (Figure 1): 1) Clustering-based, where data is represented in terms of pairwise similarity between the voxels; 2) Graph-based, where data is represented as a graph, where each node is a voxel and an edge between each pair of nodes captures the similarity between them; 3) Statistical, where data is represented as a single vector per voxel; 4) Surface-based, where data is represented as a single vector per cortical voxel and cortex is projected onto a two-dimensional surface. Each of these methods makes a set of assumptions about the underlying grouping structure of the data that will affect the structure of the resulting parcellation. In what follows we will briefly describe each class of methods. A more detailed description of each method is provided in Table 2.

**Clustering-based-** The clustering-based parcellation algorithms use pairwise similarity between voxels to divide them into groups that have higher within-group pairwise similarity compared to between-group pairwise similarity. In the data mining literature, each group is called a 'cluster', and the algorithm used to assign data points to different clusters is called a 'clustering algorithm'. While numerous clustering algorithms are available in data mining literature, three clustering algorithms that have been used for brain parcellation are K-means and K-medoids clustering, hierarchical clustering, and spectral clustering. Clustering-based methods require the researcher to specify the number of regions or parcellation granularity and are not capable of extracting the underlying number of groups themselves. Each method has specific strengths and drawbacks, summarized in Table 2. When applied to datasets that do not meet these assumptions, the clustering algorithm might be incapable of grouping voxels into their 'true' grouping structure. In addition to these clustering algorithms, preliminary results have shown the potential for a more recent clustering algorithm, affinity propagation (Frey and Dueck 2007), to be used for brain parcellation . One of the main advantages of affinity propagation is that it does not require the user to specify the number of clusters.

| Clustering algorithm | Brief Description | Strengths | Drawbacks |
|---|---|---|---|
| **Clustering-based** | | | |
| K-means and K-medoids (Tan, Steinbach, and Kumar 2006) | The K-means algorithm assumes voxels assigned to each cluster form a sphere around the cluster centroid. The initial choice of centroids is random. Each voxel is assigned to the cluster with the most similar centroid. Consistency of the results must be compared across | i) The algorithm is fast. ii) A variant of the algorithm, known as fuzzy | i) Biased towards equal size and globular shaped groups (Tan, Steinbach, and Kumar 2006). ii) Sensitive to the initial |
| | several random initial centroids to determine if the clustering is reliable. If the algorithm does not consistently converge to the same solution, the existence of clusters can be called into question. A variant of the method called K-medoids is less sensitive to outliers and has been for parcellation. Both k-means and k-medoids algorithms have been used for functional parcellation of the brain (Bellec et al. 2010; Lee et al. 2012; Thirion et al. 2014; Najafi et al. 2016; Salehi et al. 2017). | K-means supports soft assignment (Lee et al. 2012). | choice of centroids.<br>iii) K-means is sensitive to outliers.<br>iii) Not suitable for regions of non-convex shapes. |
| Hierarchical Clustering (Johnson 1967) | The hierarchical clustering algorithm produces multiple clustering solutions at different granularities in a nested hierarchical arrangement. Granularity ranges from assigning each voxel to a single cluster to assigning all voxels to the same cluster. Bigger clusters are constructed by merging the most similar smaller clusters. An extension to the algorithm has been proposed that merges smaller groups only if their merger has a lower variance than a random merge of groups to control for false-positive discovery rate (Wong et al. 2016). Hierarchical clustering has been used for functional (Thirion et al. 2014; Moghimi, Lim, and Netoff 2017; Arslan and Rueckert 2015) and structural (Moreno-Dominguez, Anwander, and Knösche 2014; Gallardo et al. 2017) parcellation of the brain. | i) Provides a range of parcellation granularities.<br>ii) Can handle data with groups of unequal size.<br>iii) Does not assume any shape for the clusters. | i) Computationally expensive (Tan, Steinbach, and Kumar 2006). |
| Spectral Clustering (Luxburg 2007) | The spectral clustering algorithm transforms the dataset into a lower-dimensional space where the grouping of the data is more apparent, then applies a clustering algorithm to divide the dataset into multiple groups (Luxburg 2007; Chung 1997). Spectral clustering has been used for functional and structural parcellation of the brain. | i) Does not assume any shape for the clusters. | i) Biased towards equal-size groups. |
| **Graph-based** | | | |
| Community detection (Fortunato and Hric 2016) | Community detection algorithms search for subsets or 'communities' of nodes in graphs that are more strongly connected to other nodes in the community compared to nodes not included in the community. Such communities or modules provide a partitioning of the graph. Community detection algorithms can estimate the optimal number of communities (e.g. (Newman 2006; Blondel et al. 2008)). Community detection has been used only for functional parcellation of the brain (Mumford et al. 2010; Power et al. 2011; Laumann et al. 2015; Najafi et al. 2016; Gordon, Laumann, Adeyemo, Gilmore, et al. 2017; Gordon, Laumann, Adeyemo, and Petersen 2017). | i) Able to estimate the number of regions that best describe the data.<br>ii) Supports soft assignment.<br>iii) Can handle data with groups of unequal size.<br>iv) Does not assume any shape for the groups. | i) Sensitive to random fluctuations in edge patterns (Fortunato and Hric 2016). To alleviate this problem, multiple samples of the same graph should be partitioned and compared. For example, different fMRI datasets from the same subject can be parcellated and compared. |
| Graph cut (Boykov, Veksler, and Zabih 2001) | The graph cut algorithm formulates the parcellation problem as an energy optimization problem. An energy function is defined by the researcher that captures the parcellation objective, which is maximizing the similarity of voxels grouped together. Other desired properties for the parcellation can be achieved by incorporating appropriate terms into the energy function. For example, an energy function can have two components, one that enforces grouping of similar | i) The energy function enables the researcher to enforce desired properties on the parcellation result.<br>ii) Able to estimate the | i) Is not guaranteed to find the optimal solution (Boykov, Veksler, and Zabih 2001).<br>ii) Weighting different components in the objective function are arbitrary and have to be decided by the researcher. |

| | voxels together, and one that enforces spatial contiguity of regions. Different terms can be weighted differently based on their relative importance. The graph cut algorithm discovers a partitioning of the graph that minimizes the energy function. Graph cut has been used for functional and structural parcellation of the brain. | number of regions that best describe the data. iii) Can handle data with groups of unequal size. iv) Does not assume any shape for the groups. | |
|---|---|---|---|
| **Statistical** | | | |
| von Mises Fisher (vMF) mixture model (Banerjee et al. 2005) | The vMF mixture model assumes that the connectivity pattern of voxels is generated from a set of vMF distributions. Each distribution has different parameters. To parcellate the brain, parameters of each distribution are estimated, and voxels generated from the same distribution are assigned to the same region. VMF distribution describes the stochastic behavior of vectors and is suitable for modeling connectivity profile vectors. The vMF mixture models have been used only for functional parcellation of the brain (Yeo et al. 2011; D. Wang et al. 2015; Kong et al. 2019). | i) Supports soft assignment. ii) The hierarchical formulation of vMF mixture models automatically establishes concordance between subject-level parcellations (Kong et al. 2019). | i) The similarity between connectivity profiles is quantified using the Pearson correlation coefficient which is not sensitive to the amplitude of the connectivity profiles, which is relevant information or parcellation. ii) Parameter estimation can be sensitive to the initial choice of parameters which is chosen randomly (Bishop 2006). iii) Not suitable if connectivity profile vectors do not form convex groups (Mitra, Pal, and Siddiqi 2003). |
| Dirichlet mixture model (DMM) (Ferguson 1973) | The Dirichlet mixture models (DMM) are a class of general mixture models that estimate the number of groups in the dataset by using a Dirichlet process. A Dirichlet process is a stochastic process, samples of which are probability distributions themselves. A sample distribution generates a mixture model, describing the number of groups in the dataset as well as how parameters of the mixture model are generated. The inference procedure estimates the parameters of the Dirichlet Process from the data. Therefore, the number of regions is learned from the data itself and does not need to be specified by the user. The DMM have been used for functional (Yeo et al. 2014; Baldassano, Beck, and Fei-Fei 2015) and structural (Baldassano, Beck, and Fei-Fei 2015) parcellation of the brain. | i) Able to estimate the number of regions that best describe the data. ii) Supports soft assignment. Iii) Hierarchical formulation of DMMs automatically establishes concordance between subject-level parcellations (S. Jbabdi, Woolrich, and Behrens 2009). | i) The family of probability distributions generated by the Dirichlet process has to be specified by the user. ii) Parameter estimation can be sensitive to the initial choice of parameters which is chosen randomly . iii) Not suitable if connectivity profile vectors do not form convex groups (Mitra, Pal, and Siddiqi 2003). |
| Dictionary Learning (Kreutz-Delgado et al. 2003) | The goal of dictionary learning algorithms is to find a dictionary with a finite number of basis functions that provide a 'sparse' representation of the data. Dictionary learning assumes that each data point is a linear combination of the basis functions. In the context of brain parcellation, dictionary learning algorithms are only applicable to fMRI data. They seek to learn a set of basis functions as well as how they are combined to generate each time-series. The algorithm estimates the weight assigned to each basis function to generate each | i) Supports soft assignment. | i) Can be sensitive to outliers (Gribonval, Jenatton, and Bach 2014). ii) Is not applicable to structural parcellation. |
| | voxel's time-series. To construct regions, voxels who are assigned the highest weight for each basis function are grouped together (Abraham et al. 2013), although more complex methods for region extraction have been proposed (Abraham et al. 2014). Independent component analysis (ICA) is a special case of dictionary learning that seeks basis functions that are independent in the sense that the weight assigned to one of the basis functions for a specific time-series has no effect on the weights assigned to other functions for the same time-series (Kreutz-Delgado et al. 2003). Probabilistic functional mode (PFM) is another special case of dictionary learning where the basis functions are identified under constraints imposed by inter-subject variability and the hemodynamic response function modeled in a probabilistic framework (Harrison et al. 2015). General formulation of dictionary learning (Varoquaux et al. 2011; Abraham et al. 2013), ICA (e.g. (Beckmann and Smith 2004; Beckmann and Smith 2005; Calhoun, Kiehl, and Pearlson 2008)), and PFM (Harrison et al. 2015) have all been used for functional parcellation of the brain. | | |
| **Surface-Based** | | | |
| Edge detection (Cohen et al. 2008) | Edge detection algorithm operates based on the observation that similarity in connectivity between an arbitrary seed voxel and all other voxels does not change smoothly and has sharp 'transition zones' over the cortical surface (Cohen et al. 2008). The position of these abrupt changes is extracted using image processing techniques and used to delineate the regions. This method has been used for functional parcellation of the cortical surface (Cohen et al. 2008; Gordon et al. 2014; Gordon, Laumann, Gilmore, et al. 2017; Gordon, Laumann, Adeyemo, and Petersen 2017; Gordon, Laumann, Adeyemo, Gilmore, et al. 2017). | i) Able to estimate the number of regions that best describe the data. ii) Supports soft assignment. iii) Can handle data with groups of unequal size. iv) Does not assume any shape for the groups. | i) The granularity of the parcellation cannot be directly controlled by the user. It is done by changing the parameters of the image processing algorithm. |
| Region growing (Blumensath et al. 2013) | Region growing algorithm was originally developed for functional parcellation of the brain but it is also applicable for structural parcellation. Region growing algorithm first constructs a two-dimensional stability map that quantifies local homogeneity of functional activity across the cortical surface. Locations with the highest homogeneity values are chosen as putative centers of the regions. Each region is then "grown" by assigning the most similar neighboring voxels to it. These regions are then grouped together using a clustering algorithm to construct bigger and final regions (Blumensath et al. 2013). Construction of the stability map requires the application of spatial smoothing which reduces the spatial precision of the parcellation result. To address this issue, a modification of this method has been proposed which calculates the gradient of homogeneity values across the stability map without smoothing it. It places the border between regions where the gradient is steepest (Dias et al. 2015). | i) Does not assume any shape for the groups. | i) In the original formulation, spatial smoothing of the map limits the spatial resolution of the results. ii) Requires a grouping algorithm to construct the final regions and suffers from any potential shortcomings of that algorithm. |

**Table 2. Summary of algorithms used for the construction of functional and structural parcellations.** A brief description of each algorithm and its strengths and drawbacks are listed. 'Soft assignment' here refers to assigning

each voxel to all regions but with different weights or degrees of membership. Soft assignment is in contrast with hard assignment where each voxel is assigned to one region only.

**Graph-based-** A graph is an abstract representation of elements of a system and their interactions. A graph consists of a set of nodes (vertices) interconnected by edges (links). In the context of brain parcellation, each voxel constitutes one node and each edge captures the similarity between the nodes. The graph can be binarized by setting the value of an edge between a pair of nodes to 1 if the similarity between them is above a selected threshold and 0, otherwise. Modeling the brain as a graph of interconnected voxels makes it possible to perform parcellation using a variety of graph partitioning algorithms that divide graphs into several subgraphs, where voxels belonging to each subgraph form a region. Several graph partitioning algorithms exist in the graph theory literature (Bichot and Siarry 2011). In the context of brain parcellation, two well-known graph partitioning algorithms have been used: Community detection, and graph cut (Table 2). The rich body of graph partitioning literature includes other powerful graph partitioning algorithms that seem suitable but have not been used for brain parcellation. One such method is multi-level graph partitioning (Karypis and Kumar 1998a, b) which is particularly suitable for partitioning large graphs such as graph models of the brain. Another is hypergraph partitioning where an edge can join any number of nodes rather than pairs of nodes (Bichot and Siarry 2011) that can be useful to study temporal changes in the brain structure over time for example during learning (Bassett et al. 2014). Temporal dynamics of brain connectivity can also be modeled using temporal networks which are graphs whose links are active only at certain points in time. Application of these modeling and partitioning algorithms to the problem of brain parcellation is yet to be explored.

**Statistical-** The statistical class of approaches assumes a statistical model for the structure of the data. The purpose of the algorithm is to estimate the parameters of the model. More specifically, each voxel's time-series or connectivity profile is assumed to be generated according to a parametric model of unknown parameter values. Voxels with similar parameter values are grouped to form regions. Statistical parcellation methods flexibly incorporate prior assumptions about the data into closed-form mathematical models. The drawback of statistical methods is that if the assumption they make is not a good fit, it will lead to spurious results. Statistical methods used for brain parcellation include the von Mises Fisher (vMF) mixture model, Dirichlet mixture model, and Dictionary learning which is a more general case of the independent component analysis (Table 2).

**Surface-based**- Surface-based class of methods operates on the surface of the cerebral cortex, treating it as a two-dimensional map. They rely on the spatial relationship between voxels as well as their similarity and are the only class of methods developed by the neuroscience community specifically for parcellation of the cortical surface. Surface-based methods capitalize on the fact that adjacent voxels on the cortical sheet have a similar functional activity or connectivity profiles except at borders between regions. Structural parcellation of the brain using surface-based methods, even though plausible in theory, has not been explored. Surface-based methods used for brain parcellation are edge detection and region growing (Table 2).

Parcellation algorithms are typically directly applied to single voxels. A few studies, however, have used small contiguous groups of voxels instead of single voxels as the elements that are grouped to form regions (Blumensath et al. 2013; Arslan and Rueckert 2015; Arslan, Parisot, and Rueckert 2015; J. Wang, Hao, and Wang 2018). These small groups consist of highly homogenous voxels and are also called supervertices (Arslan and Rueckert 2015). The number of superverices is typically in the order of thousands to hundreds compared to tens of thousands of single voxels. The property of interest for each super vertex is calculated from its constituent voxels. For example, the functional activity of a supervertex has been calculated as the average time-series of its voxels (Blumensath et al. 2013; Arslan and Rueckert 2015). Using supervertices instead of single voxels has three main advantages. First, it increases the signal to noise ratio by combining data from multiple highly similar voxels. Second, it reduces the computational cost of parcellation by reducing the size of the dataset. Third, supervertices facilitate comparison of parcellation results across subjects . Functional and structural parcellations, unlike anatomical parcellations, do not provide a unique definition or label for each region. When comparing parcellations of two subject brains, it is not clear which region in one subject corresponds to which region in the other. Due to inter-individual differences, spatial location and size of regions vary across parcellations of subject brains. Image registration prior to parcellation only provides a correspondence between single voxels. Supervertices, on the other hand, represent small regions in the brain with homogenous functional or structural properties whose spatial location and extent are similar across subjects but do not exactly match. Correspondence between regions at supervertex level can be established by grouping supervertices with maximal spatial overlap across subjects. Incorporation of supervertices into any of the parcellation algorithms covered in this manuscript is feasible and straightforward.

Each class of parcellation algorithms has specific assumptions about the underlying structure of the data which dictates how the voxels are grouped to form regions. These assumptions are not mutually exclusive. Different methods capitalize on different and potentially complementary characteristics of the dataset to parcellate the brain. Therefore, formulating the parcellation procedure as an optimization problem that combines the objectives from several parcellation algorithms has the potential to outperform each of the individual algorithms. In fact, one example of such a combinatorial approach, combining edge detection with vMF mixture model, has been reported to outperform the edge detection algorithm when used on its own (Schaefer et al. 2018).

Some parcellation algorithms such as K-means and von Mises-Fished mixture model require initialization. For example, the K-means algorithm requires an initial set of cluster centroids. Different initializations produce different parcellations because these algorithms are not guaranteed to converge to the optimal parcellation solution (Ryali et al. 2013). The question then is how to construct a single parcellation given the variability introduced by the sensitivity of the algorithm to initialization. One approach is to choose the initial conditions that meet a certain criterion hypothesized to result in better parcellations. For example, instead of choosing random initial centroids for the K-means algorithm, they can be set to voxels that maximally summarize the dataset across multiple subjects to reduce the sensitivity of the algorithm to outlier data points (Salehi et al. 2017). Another approach is to generate a set of parcellations from many randomly chosen initializations and pick the "best parcellation", where the definition of the "best parcellation" depends on the criterion set forth by the algorithm (Yeo et al. 2011; Jin et al. 2015). For example in the case of the von Mises-Fisher mixture model the "best parcellation" across different initializations has been defined as the parcellation with the highest likelihood (Yeo et al. 2011). Both approaches seek to find initializations that produce better parcellation results. But neither approach resolves the issue of sensitivity of the parcellation result to initialization. A third approach that addresses this problem is consensus learning. Consensus learning alleviates sensitivity to initialization by producing many parcellations from different initializations and combining them to construct an aggregate parcellation that captures their commonalities (Strehl and Ghosh 2002). The aggregate parcellation is constructed by grouping voxels with similar membership patterns across the individual parcellations (Ryali et al. 2015). The underlying assumption behind consensus learning is that patterns that are shared across different initializations are more likely to reflect the grouping structure of the data rather than random variations introduced by initializations (Strehl and Ghosh 2002). Consensus learning is a powerful tool commonly used by data scientists to increase robustness to initialization and is a more reliable method for constructing a parcellation that is faithful to the underlying structure of the dataset than the other approaches.

## 3.3 Level of analysis

Functional and structural parcellations have been constructed at both subject and group levels. Subject-level parcellations are constructed by applying the parcellation algorithm to an individual's dataset (Blumensath et al. 2013; Moreno-Dominguez, Anwander, and Knösche 2014; Laumann et al. 2015; Salehi et al. 2017; Gordon, Laumann, Gilmore, et al. 2017; Braga and Buckner 2017; Kong et al. 2019). Group-level parcellations are constructed using composite datasets from a group of individuals .

Each level of analysis has its pros and cons. The first generation of functional and structural parcellations was constructed at group level to boost the signal to noise ratio (Braga and Buckner 2017). This is especially true for functional parcellations constructed using resting-state data that typically lasted about 5-10 minutes in duration for each individual, which is not adequate for the construction of reliable functional parcellations (Laumann et al. 2015), although a recent study has shown more complex statistical models are capable of constructing reliable individual-level parcellations with such short fMRI datasets (Kong et al. 2019). As a result, combining individual datasets used to be the standard solution to reduce noise. In addition, group-level parcellations capture the common organization of the brain shared by the group and reveal the structure of an "average brain" and not idiosyncrasies unique to any individual brain. Construction of group-level parcellations also avoids the concordance problem where the correspondence between regions of individual parcellations is not known.

When constructing group-level parcellations, it is preferable to use subject cohorts that are close in age or divide the subjects into multiple age groups and parcellate each group separately. The reason is that anatomy, functional organization, and white matter integrity of the brain changes across the human lifespan (Hedden and Gabrieli 2004). At least for functional parcellations, it has been shown that organization and properties of the parcellation differ for different age groups (Honnorat et al. 2015; Han et al. 2018), coinciding with changes in cortical thickness and morphology (Han et al. 2018). As a result, combining datasets from a cohort with highly variant ages to construct group parcellations would result in blurring of the boundaries between regions as the boundaries in different age groups might not align. Typical age groups span between 6 (Honnorat et al. 2015) to 14 years (Han et al. 2018). How and to what degree structural parcellations change across the human life span has not yet been studied, but given the observed alterations of white matter tracts in the process of aging (Liu et al. 2017), it is important to also characterize these changes in terms of parcellation properties and to separate subjects based on their age for construction of group-level structural parcellations.

The main drawback of parcellating the brain at the group level is that the unique features of brain organization specific to each individual are averaged out in the parcellation process. Recent studies have shown that reliable functional parcellation at the individual level is possible using longer duration fMRI datasets (typically more than 30 minutes (Laumann et al. 2015; Gordon, Laumann, Gilmore, et al. 2017)). These studies have shown that organization of individual brains have specific characteristics that have not been observed in group-level parcellations (Laumann et al. 2015; Braga and Buckner 2017; Glasser et al. 2016; Gordon, Laumann, Adeyemo, Gilmore, et al. 2017; Gordon, Laumann, Adeyemo, and Petersen 2017; Gordon, Laumann, Gilmore, et al. 2017). For example, individual-level parcellation of the brain fractionated the default mode network estimated at the group level into two separate networks (Braga and Buckner 2017), showing that parcellating the brain at a group level can obscure some features of the brain organization that are only observed when the brain is parcellated at an individual level, possibly due to inter-individual variability in spatial location of smaller regions (Laumann et al. 2015; Braga and Buckner 2017; Gordon, Laumann, Gilmore, et al. 2017). A similar comparison across individual and group-level structural parcellations is yet to be done. We expect the results to be similar to what was observed with functional parcellations, where features present at individual-level parcellations are absent at the group level.

The construction of reliable individual-level parcellations allows the advancement of different lines of research. One relevant area of research is linking brain organization to behavior by identifying which characteristics of individual-level parcellations are correlated with and potentially underlie behavioral inter-individual differences (Finn et al. 2015). For example, one group has shown that some features of the brain parcellation at the individual level are significantly albeit mildly correlated with performance in complex cognitive tasks (Kong et al. 2019). Scrutinizing which characteristics of the brain organization are better predictors of individuals' behavior provides researchers with viable candidates for neural mechanisms that underlie them. Another venue where individual-level parcellations can be very beneficial is biomarker discovery. Comparing individual-level parcellations of healthy individuals with that of patients might reveal more subtle changes in brain organization that are linked to the disease but are not detectable at group-level parcellations. Furthermore, individual-level parcellations can also reveal the extent to which those pathological changes correlate with the severity of symptoms and can guide researchers in designing informed and targeted treatments (Jeste and Geschwind 2014). Such analysis is especially useful for studying complex mental disorders with highly heterogeneous patient populations such as schizophrenia (Takahashi 2013), substance abuse (Litten et al. 2015; Li and Burmeister 2009), and autism (Jeste and Geschwind 2014).

The choice of level of analysis ultimately depends on the specific application of the parcellation. When unique features of the individual brain are irrelevant to the study or when the available data from each individual is not sufficient for constructing reliable individual-level parcellations (for example short fMRI datasets) group-level parcellations can be constructed.

There are several approaches for combining datasets from a cohort of subjects to construct a group-level parcellation. The first approach, referred to here as the 'ante-hoc' approach, is to combine the individual datasets prior to applying the parcellation algorithm. Datasets have been combined by averaging pairwise similarity between voxels across all subjects (Yeo et al. 2011; Craddock et al. 2012; Gordon et al. 2014; Andrew James, Hazaroglu, and Bush 2015; Laumann et al. 2015; Gallardo et al. 2017). In the case of functional parcellations, an alternative is to concatenate the time-series across subjects before calculating the pairwise similarity between voxels (Honnorat et al. 2015; Moghimi, Lim, and Netoff 2017; Schaefer et al. 2018).

Another approach, referred to here as the 'post-hoc' approach, is to parcellate each individual dataset separately and combine the resulting parcellations by employing consensus learning (van den Heuvel, Mandl, and Hulshoff Pol 2008; Bellec et al. 2010; Craddock et al. 2012; Arslan and Rueckert 2015; Ryali et al. 2015). Consensus learning groups voxels based on their co-assignment frequency across individual-level parcellations. More specifically, voxels that are frequently grouped together at the individual level, are also grouped together at the group level. Hence, consensus learning captures the common structure in individual-level parcellations. A comparison between the 'ante-hoc' and 'post-hoc' approaches has shown that the 'post-hoc' approach results in group-level parcellations that are in higher agreement with individual-level parcellations (Craddock et al. 2012) and are more reproducible across groups (Arslan et al. 2017).

Finally, a third approach is to enforce constraints on the structure of every individual-level parcellation such that it is highly similar to the group, i.e. construct "group consistent" individual-level parcellations, in addition to a group-level parcellation that maximally captures commonalities between the individual-level parcellations (Varoquaux et al. 2011; Abraham et al. 2013; Shen et al. 2013; Wang et al.

2015; Arslan, Parisot, and Rueckert 2015; Parisot et al. 2015; Parisot, Arslan, et al. 2016; Gordon, Laumann, Adeyemo, Gilmore, et al. 2017; Chong et al. 2017; Zhao, Tang, and Nie 2019). We refer to this approach as the 'simul-hoc' approach. The advantage of the simul-hoc approach is that it produces both individual- and group-level parcellations. In addition, concordance between regions of the individual-level parcellations is automatically established by the algorithm. However, it is not yet clear how much of the unique features of each individual is preserved by this method as individual-level parcellations are constrained to be as similar as possible to the group-level parcellation.

Construction of group-level parcellations requires registration of the individual brains to a common stereotaxic space so that a one to one correspondence between single voxels exists across subjects. The standard registration method is to use macro-anatomical landmarks of the brain, i.e. brain morphology, to align individual images (Langs et al. 2016). The implicit assumption of this method is that corresponding voxels across the aligned images have similar functional roles and consequently similar properties such as functional activity or connectivity profiles. However, the relative position of voxels with respect to anatomical landmarks of the brain is not perfectly correlated with their connectivity profiles (Mueller et al. 2013; Langs et al. 2016). The result is reduced specificity of group-level parcellations that rely on morphological alignment, since corresponding voxels across individuals are assumed to have similar connectivity profiles, when in fact they do not. This is especially the case in the associative cortex where the correlation between morphology and connectivity profile is weaker (Langs et al. 2016). To resolve this problem, a recent study proposed aligning individual brains based on the functional connectivity profile of voxels and not their spatial location (Langs et al. 2016). More specifically, correspondence between voxels across different subjects is established based on the similarity between the voxels' functional connectivity profiles. This approach has resulted in more reproducible parcellations compared to parcellations constructed using morphological alignment (Langs et al. 2016). The increase in specificity promised by this method of registration is readily applicable to any parcellation algorithm. This approach is also applicable to the construction of group-level structural parcellations, where individual brains can be aligned based on their structural connectivity profiles and then parcellated.

## 3.4 Additional remarks

Functional parcellations

FMRI datasets are preprocessed before being used for parcellation. The preprocessing pipeline is generally consistent across parcellation studies and includes a correction for slice timing and instability noise, artifact removal, temporal bandpass filtering, registration to a common stereotaxic space, and spatial smoothing . However, different functional parcellation studies differ mostly in two parameters: the choice of temporal filter cut off frequencies and the extent of spatial smoothing. These parameters are often reported without any justification as to how they were chosen. The sensitivity of parcellation results to the specific choice of preprocessing parameters has not been systematically studied. We argue that functional parcellation studies should report on the sensitivity of their results to the specific choice of preprocessing parameters. If the results are sensitive to any of the parameters, values that produce optimal parcellation results should also be reported.

To the best of our knowledge, the impact of cut off frequencies on parcellation results has not been reported for any of the available parcellation algorithms, which is concerning because the choice of the cutoff frequencies has been shown to impact the test-retest reliability of fMRI datasets (Shirer et al. 2015) and can potentially improve the reproducibility of functional parcellations as well.

Spatial smoothing has been reported to have a negligible impact on parcellation results for a handful of algorithms: vMF mixture model (Yeo et al. 2011), community detection algorithm (Goulas, Uylings, and Stiers 2012), and spectral clustering algorithm (Kelly et al. 2010; Shen et al. 2013). It is imprudent however to assume that all parcellation algorithms and granularities are resilient to the extent of spatial smoothing without systematically testing them. An alternative solution is to measure the "cluster tendency" of fMRI datasets at different spatial smoothing scales. Cluster tendency quantifies to what degree any grouping structure exists in the data and can be quantified using measures such as the Hopkins statistics (Tan, Steinbach, and Kumar 2006). The interested reader is referred to (Adolfsson, Ackerman, and Brownstein 2018) for a review of other methods of measuring cluster tendency. The advantage of using this approach to comparing parcellation results at different smoothing scales is that it is computationally less expensive and is independent of the parcellation algorithm and granularity.

In addition to the preprocessing pipeline described in the previous paragraph, some studies have removed temporal autocorrelation from fMRI time-series, a process known as prewhitening (Woolrich et al. 2001; Christova et al. 2011). Prewhitening removes spuriously high cross-correlations that are due to temporal autocorrelation and do not reflect shared neural activity between voxels. Prewhitening has been shown to be a promising preprocessing step for biomarker discovery and estimation of task activated voxels (Olszowy et al. 2019). One study has reported that prewhitening fMRI time-series changes the structure of functional parcellations by as much as 10% and improves the quality of the parcellation (Moghimi, Lim, and Netoff 2017). Collectively, these results point at prewhitening as a promising preprocessing step for improving functional parcellations.

The effective spatial resolution of functional parcellation is limited by the spatial autocorrelation inherent to fMRI datasets (Eklund, Nichols, and Knutsson 2016). Spatial autocorrelation is the tendency of nearby voxels to share similar time-series and hence higher correlation coefficients. One cause of spatial autocorrelation is head motion, despite performing motion estimation and regression on the fMRI data (Power et al. 2012). As a result, high correlation values between nearby voxels do not necessarily mean that they share the same functional role and is therefore reflective of both biological and artifactual signals. This poses a challenge for the construction of functional parcellations (Power et al. 2011). Power and colleagues proposed setting the correlation coefficients between voxels within a certain distance of each other to zero (Power et al. 2011). This correction resulted in the removal of slightly more than 4% of pairwise positive correlation values. Until further advancements allow for more effective motion artifact removal, this technique is a simple and effective method to reduce the impact of non-biological spurious correlations.

Signal to noise ratio of fMRI datasets is not homogenous across the brain (Yeo et al. 2011; Welvaert and Rosseel 2013; Yeo et al. 2014). Consequently, the reliability of the parcellation results cannot be assumed to be homogenous either (Yeo et al. 2014). The cerebellum, subcortical structures, and portions of the limbic system are reported to be regions with high susceptibility to imaging noise (Yeo et al. 2014). One plausible remedy to alleviate the impact of imaging noise is to use similarity between functional connectivity profiles instead of functional activity as the similarity measure (Yeo et al. 2011). In the absence of grand truth, it is not known how effective this method is. When using a functional parcellation to study any of these regions, one must be aware of potential inaccuracies in the parcellation process.

A fundamental yet implicit assumption made about functional parcellations is that there exists a stationary functional parcellation of the brain and the goal of the parcellation algorithm is to discover that parcellation using the functional activity of the brain. This assumption is challenged by the observation that functional connectivity profiles are not stationary during resting-state (Smith et al. 2012; Allen et al. 2014). Moreover, connectivity profiles do not change randomly over time but switch between several recurring patterns or brain states (Allen et al. 2014). Switching between states happens on a time scale of tens of seconds. Several methods have been used to identify different brain states. One study performed the temporal independent component analysis to identify 21 brain states . Several other studies have constructed functional connectivity profiles within sliding time windows and applied clustering algorithms to group similar profiles to form brain states . These studies have typically identified 7-9 brain states. The non-stationary nature of functional activity has important implications for functional parcellation. A few studies have shown that the number of regions or the location of region boundaries changes as a result of brain state . Merging parcellations constructed at different states produces a parcellation that aligns better with histologically defined regions compared to a single parcellation constructed using the entire duration of functional activity (Chen et al. 2016; Ji et al. 2016). While merging parcellations constructed from different brain states is one way of dealing with the dynamic nature of the functional organization of the brain, it is not clear whether it is the optimal strategy. Constructing a separate functional parcellation for each brain state might be more informative about the organization of the brain at every point in time. Further research is required to explore whether using state-specific functional parcellations is beneficial for studies that use these parcellations. For example, it is possible that the difference in brain organization between two groups of interest, e.g. a cohort of subjects with a neurological disorder and a control group, is emphasized during one brain state and not the other. Using state-specific parcellations can highlight these differences that would otherwise go unnoticed.

Given the dynamic nature of fMRI datasets, one important question is whether and how functional parcellations constructed from resting-state and task-evoked fMRI activity might differ. At the group level, it has been reported that resting-state functional parcellation is highly similar to task-evoked functional parcellation . However, recent studies have shown that the spatio-temporal structure of fMRI activity captured in resting-state is different from task-evoked activity . Furthermore, several recent studies by Salehi and colleagues have shown that at the subject level, the organization of functional parcellations changes as a function of the type of fMRI data used for their construction, i.e. resting-state vs task-evoked (Salehi, Greene, et al. 2019; Salehi, Karbasi, et al. 2019). In addition, the organization of the brain depends on the task the subject is engaged in. Salehi and colleagues constructed several functional parcellations using task-evoked fMRI activity collected during a battery of cognitive tasks in addition to several resting-state sessions. They observed that a functional parcellation constructed from the evoked activity during one task would differ from a parcellation constructed from the evoked activity during another task (Salehi, Greene, et al. 2019; Salehi, Karbasi, et al. 2019). Interestingly, the extent of task-related reconfiguration depended on how well the subject was engaged in the task, i.e. subject's performance (Salehi, Greene, et al. 2019). Regions with the highest reconfiguration across brain states were also more variable across subjects (Salehi, Karbasi, et al. 2019). These findings again raise the question of how to construct a relevant functional parcellation. Similar to state-dependent functional parcellations, task-dependent functional parcellations can also be merged to construct a single parcellation. The structure of such a consensus parcellation and its commonalities and differences with brain state-specific parcellations are interesting topics that remain unexplored. Alternatively, the construction and use of task-specific functional parcellations need to be investigated. If the parcellation is intended for studying task-evoked activity, it is appropriate to use a separate fMRI dataset collected during the same task to construct a functional parcellation that is most relevant for that task.

In addition to the algorithms discussed in this section, functional parcellations constructed from task-evoked fMRI activity have also been constructed using meta-analytic approaches that group voxels based on their co-activation patterns across several tasks as reported across a large number of studies. Of note are two approaches known as meta-analytic connectivity modeling (MACM) (Eickhoff et al. 2011) and meta-analytic activation modeling-based parcellation (MAMP) (Yang et al. 2016). These approaches group voxels that are modulated similarly across different tasks. The advantage of the meta-analytic approach is that it enables developing hypotheses about the functional roles of the resulting regions based on the tasks that maximally activated them (Eickhoff et al. 2011). However, combining datasets that were collected by different groups using different scanners and acquisition protocols causes spatial blurring and reduces the precision of the parcellation result (Eickhoff, Yeo, and Genon 2018). In addition, the meta-analytic approach is not suitable for whole-brain parcellation. To be able to construct a whole-brain parcellation, a set of tasks that can modulate all the voxels in the brain at least once are required. Such an approach is not feasible as tasks for reliably activating some locations in the brain are not yet known (Klein et al. 2007). As a result, the meta-analytic approach to functional parcellation has been limited to parcellation of parts of the brain that are reliably modulated by known tasks, such as the amygdala (Bzdok et al. 2013) and dorsolateral prefrontal cortex (Cieslik et al. 2013).

Structural parcellations

DWI datasets are preprocessed prior to tractography and application of the parcellation algorithm. Preprocessing of diffusion signal includes intensity normalization to remove the effect of magnetic field inhomogeneity, removal of artifacts caused by head motion and eddy currents, and correction of the dataset for gradient nonlinearities (Glasser et al. 2013; Sotiropoulos et al. 2013). The preprocessed signal is then used for tractography. Tractography methods vary in three aspects: i) deterministic vs. probabilistic tractography; ii) Single direction vs. multi directions tractography; iii) Choice of seed voxels for tractography. Structural parcellation studies have typically used probabilistic tractography, with a few exceptions that used the deterministic approach (e.g. (Roca et al. 2009, 2010)). Both single-directional (e.g. (Rushworth, Behrens, and Johansen-Berg 2006; Devlin et al. 2006; Klein et al. 2007)) and multi-directional (e.g. (Beckmann, Johansen-Berg, and Rushworth 2009; Mars et al. 2011; Baldassano, Beck, and Fei-Fei 2015)) tractography algorithms have been used for the construction of structural parcellations. The choice of seed voxels for tractography also varies among different parcellation studies. While multiple studies have used all gray matter voxels as seed locations, it has been argued that since the penetration of white matter into the gray matter is not deep in the cortex, gray matter voxels located at the border between gray and white matter are a better candidate for seed locations (Anwander et al. 2007), and this approach has been adopted for several structural parcellations (Anwander et al. 2007; Mars et al. 2011; Schubotz et al. 2010; Sallet et al. 2013). The sensitivity of parcellation results themselves to none of these factors has been studied yet, however. Without a systematic study of the impact of different tractography algorithms, it is not possible to evaluate structural parcellation algorithms effectively, as any inaccuracies in the parcellation result can be caused by the tractography algorithm and not the parcellation algorithm.

The precision of structural parcellations is contingent upon accurate estimation of white matter fibers and is therefore limited by the precision of the tractography process. Despite the recent developments in enhancing tractography algorithms (e.g. ), recovering complex fiber patterns such as crossing or fanning white matter fiber bundles using the existing DWI technology and tractography algorithms has been a challenge (Bastiani et al. 2012). In addition, tractography techniques are not capable of detecting intra-areal and u-fibers (fibers connecting adjacent gyri) (Anwander et al. 2007), have a bias towards terminating the fibers in the gyral surface rather than sulcal banks (Reveley et al. 2015; Donahue et al. 2016; Schilling et al. 2018), and are more likely to identify shorter fibers (Liptrot, Sidaros, and Dyrby 2014). The current state-of-the-art tractography techniques also have a high false-positive rate. It has been estimated that more than 30% of the identified connections by tractography algorithms do not exist (Maier-Hein et al. 2016). A comparison of tractography results with results of invasive axonal tracing in macaque monkeys has shown that tractography methods that are able to detect more white matter pathways confirmed by anatomical tracing (high sensitivity), also result in more false-positive pathways that were not observed in the tracing study (low specificity) (Thomas et al. 2014; Donahue et al. 2016). Collectively, these limitations indicate that the current tractography technology is not capable of accurately recovering all the white matter fibers in the brain.

## 4. Discussion

In this paper, we reviewed different brain parcellation methods and modalities, discussed their relative strengths and weaknesses, and described various approaches for evaluating them. In what follows, we will cover further considerations regarding different imaging modalities, preprocessing of MRI data, and parcellation methodology. We will then discuss future directions.

Different modalities of MRI data capture different types of information and have different characteristics. Each imaging modality poses different challenges to the parcellation process. The main challenge for the construction of anatomical parcellations is inter-subject variabilities which make image registration challenging. Functional parcellations use functional activity that is spatially autocorrelated and non-stationary (Laumann et al. 2015). Structural parcellations are constructed using white matter trajectories that are challenging to estimate.

Functional and structural parcellations have three main differences with anatomical parcellations. First, regions in functional and structural parcellations constitute of voxels that share similar biological properties in contrast to anatomical parcellation where voxels are grouped based on their spatial position with respect to macro-anatomical landmarks. Second, the automated data-driven approaches used for functional and structural parcellations are free from subjective biases introduced by manual involvement in constructing anatomical parcellations. Third, functional and structural parcellations are capable of parcellating the brain at different granularity levels, whereas anatomical parcellations have a fixed number of regions decided by anatomy experts.

One of the crucial preprocessing steps in brain parcellation is the registration of subject brain images to a stereotaxic space. The stereotaxic space is a typical brain image and registration is performed by aligning subject brains based on their morphology. The underlying assumption is that voxels located in the same spatial location across different subjects serve the same functional purpose. However, morphology and functional role are better correlated in sensory-motor areas but poorly correlated in the associative cortex (Fischl et al. 2008). As a result, in recent years several attempts have been made to develop registration methods that are based on properties other than the morphology of the brain, for example, alignment of functional connectivity patterns (Conroy et al. 2013; Robinson et al. 2014; Langs et al. 2016), myelin content (Robinson et al. 2014), or a combination of them (Robinson et al. 2014). The registration using these features has improved alignment between subjects and has been shown to improve the quality of parcellation results (Langs et al. 2016).

Establishing concordance between regions of different parcellations, which refers to finding the corresponding region or regions across a pair of parcellations, is necessary for comparison of subject brains. Such correspondence, by definition, exists across anatomical parcellations due to the unique definition of the sulci bordering each region. Data-driven methods used for functional and structural parcellation, on the other hand, result in the assignment of each voxel to a region without any explicit definition for the regions. Traditionally the need to establish concordance was avoided by constructing group-level parcellations instead of subject-level parcellations (e.g. (Beckmann and Smith 2005; van den Heuvel, Mandl, and Hulshoff Pol 2008; Thirion et al. 2014)). More recently, several studies have constructed group constrained subject-level parcellations that enforces a common structure among subject parcellations and automatically establishes concordance between them . As previously discussed, group-level parcellations obscure unique subject-level features of brain organization. Therefore, the facilitation of studying the brain at the subject level is necessary. Some parcellation methods automatically match regions across subject parcellations (Table 2). For clustering-based parcellation methods, it is possible to match regions across subjects using an image-cosegmentation formulation of the problem, where the same objects in multiple images are jointly segmented (e.g. ). Similarly, the same region can be delineated across different subject parcellations by introducing proper constraints to the formulation of the parcellation problem . However, these solutions are specific to clustering-based parcellation methods. In general, correspondence between a pair of parcellations is established post-hoc. One method to establish a correspondence between regions of two subject parcellations is to match regions with the highest spatial overlap (Blumensath et al. 2013; Shen et al. 2013; Moreno-Dominguez, Anwander, and Knösche 2014). The drawback of this method is that it does not have a unique solution. Also, if the two parcellations do not have the same number of regions, regions that cannot be matched are discarded. Gordon and colleagues have used the Hungrian algorithm (Bourgeois and Lassalle 1971) to establish concordance between parcellations (Gordon, Laumann, Adeyemo, and Petersen 2017). The Hungarian algorithm, although computationally expensive, is capable of many-to-one assignments (Dondeti and Emmons 1996) which allows matching regions between subject parcellations with different numbers of regions.

Contiguity of the regions produced by parcellation is a desirable property. Anatomical parcellations, by definition, partition the brain into contiguous regions. The data-driven methods used for functional and structural parcellations, in general, are not guaranteed to result in regions with a contiguous set of voxels, however. Construction of a parcellation with contiguous regions is a straightforward problem, and several simple approaches have been used to enforce contiguity of the resultant regions in functional and structural parcellations. One approach is to impose a spatial constraint on the connectivity pattern of voxels so that non-contiguous voxels are not grouped (Craddock et al. 2012; Blumensath et al. 2013; Moghimi, Lim, and Netoff 2017). Another approach is to introduce statistical priors that encourage contiguity into the parcellation problem . A third approach is to use image segmentation algorithms to extract contiguous brain regions from non-contiguous regions provided by the parcellation algorithm post-hoc (Abraham et al. 2014). Extracting contiguous regions post-hoc has the advantage of enforcing fewer constraints to the parcellation problem.

The majority of the methods discussed in this review assign each voxel strictly to a single region, which is known as 'hard assignment'. An alternative approach, known as 'soft assignment', is to assign each voxel a degree of membership to each region, allowing regions to overlap (Eickhoff et al. 2015). This approach is especially relevant when the goal of parcellation is to parcellate the brain into its functional networks rather than contiguous regions (Yeo et al. 2011, 2014). Functional networks consist of several spatially distant contiguous regions. Previous studies have shown that a brain region can be recruited by more than one network (Yeo et al. 2014, 2015; Najafi et al. 2016). In fact, more than 40% of voxels have multiple network assignments. The majority of these voxels are located in the associative cortex (Yeo et al. 2014; Najafi et al. 2016). Soft-assignment allows for the characterization of overlap between different brain networks and how it changes depending on the task the subjects are engaged in (Najafi et al. 2016). Examination of overlap patterns is instrumental in studying how different brain networks interact and which functions each contiguous region within these networks serve. Several parcellation methods are capable of soft assignment of voxels (Table 2).

Our focus in this manuscript was on brain parcellations constructed using a single imaging modality. In recent years, several attempts have been made to combine multiple imaging modalities to construct parcellations that group voxels that are similar in multiple attributes. For example, parcellations constructed using fMRI and DWI datasets consist of regions with voxels that share similar functional and structural connectivity profiles. The rationale behind multimodal parcellation is the observed correlation between different characteristics of brain tissue, i.e. brain regions with similar activity also have similar structural connection profiles and microstructure (Passingham, Stephan, and Kötter 2002). Therefore, by combining different datasets as multiple sources of information a more reliable parcellation can be constructed. Several mostly preliminary attempts to combine different modalities include combining fMRI with morphology as measured by T1-weighted images , fMRI with DWI , morphology with DWI , and fMRI with DWI and myelin maps . These studies have reported only moderate improvements in parcellation quality over monomodal parcellations. The most prominent effort in the construction of a multimodal brain parcellation has combined resting-state and task-evoked fMRI activity with cortical thickness, and myelin content of voxels (Glasser et al. 2016). A border would be delineated between two voxels if they differ from each other in at least two of the four measured properties. The most useful property was reported to be resting-state fMRI activity, followed by task-evoked fMRI activity, myelin content, and cortical thickness. The informativeness of each property depended on its location on the cortical surface. It is intuitive to expect that since this parcellation is combining information from four modalities, voxels within each region would also be homogenous in other properties not measured by the used modalities, such as structural connectivity profiles as measured by DWI. However, it does not seem to be the case. A comparison between a structural parcellation and the multimodal parcellation of Glasser and colleagues reported the total agreement between the two parcellations to be at 28% (Gallardo et al. 2017). One obvious explanation for this observation is that DWI was not used by Glasser and colleagues. However, the homogeneity of functional activity of voxels within each region is also lower in the multimodal parcellation of Glasser and colleagues compared to unimodal functional parcellations (Arslan et al. 2017), even though functional activity was used in the construction of the multimodal parcellation. It seems that the addition of other modalities results in a parcellation that is less faithful to brain organization captured by each modality, possibly because regions identified using different imaging modalities do not necessarily align. Convergence between different imaging modalities has been reported in some parts of the brain such as insula (Kelly et al. 2012), superior parietal lobule (Jiaojian Wang et al. 2015), and dorsal premotor cortex (Genon et al. 2017), but the available whole-brain multimodal parcellations have revealed that different modalities might not coalesce everywhere in the brain.

It would be useful information to know where in the brain and at what parcellation granularity different modalities diverge from one another. This information can guide us in answering another open question: how discrepancies between different imaging modalities are to be dealt with in the context of brain parcellations. For example, if two voxels have drastically different structural connectivity profiles but very similar functional activity, should they be assigned to different regions? Glasser and colleagues required at least two properties to be different for two voxels to be assigned to separate regions. Others have argued that voxels that differ in at least one property should be assigned to separate regions as different modalities capture different organizational axes of the brain (Eickhoff, Yeo, and Genon 2018). Another study has suggested assigning different weights to different modalities according to their relative reliability, for example putting less weight on structural connectivity compared to functional connectivity because tractography algorithms have a bias towards ending the tracts on the gyral surface, resulting in structural parcellations that align with cortical sulci . A detailed understanding of the pattern of convergence and divergence between different modalities is however required to decide how to deal with them. The divergence between different imaging modalities also challenges the historical concept of cortical regions where neurons within different brain regions are assumed to be distinct from each other in several properties (Eickhoff, Constable, and Yeo 2018). In light of the recent findings, a call for revising the concept of "brain region" has been put forward (Eickhoff, Constable, and Yeo 2018).

**Acknowledgements:** This work was supported by NIH grant R01 DA038984-01A1, NSF Grant IIS-1850204 and MnDRIVE Brain Conditions Postdoctoral Fellowship. The authors would like to thank Dr. Jazmin Camchong for her helpful advice.

**Declarations of interest:** none

for Cognitive State Classification." In *Graphs in Biomedical Image Analysis and Integrating Medical Imaging and Non-Imaging Modalities*, 32–42. Springer International Publishing.

Moreno-Dominguez, David, Alfred Anwander, and Thomas R. Knösche. 2014. "A Hierarchical Method for Whole-Brain Connectivity-Based Parcellation." *Human Brain Mapping* 35 (10): 5000–5025.

Mueller, Sophia, Danhong Wang, Michael D. Fox, B. T. Thomas Yeo, Jorge Sepulcre, Mert R. Sabuncu, Rebecca Shafee, Jie Lu, and Hesheng Liu. 2013. "Individual Variability in Functional Connectivity Architecture of the Human Brain." *Neuron* 77 (3): 586–95.

Mumford, Jeanette A., Steve Horvath, Michael C. Oldham, Peter Langfelder, Daniel H. Geschwind, and Russell A. Poldrack. 2010. "Detecting Network Modules in fMRI Time Series: A Weighted Network Analysis Approach." *NeuroImage* 52 (4): 1465–76.

Najafi, Mahshid, Brenton W. McMenamin, Jonathan Z. Simon, and Luiz Pessoa. 2016. "Overlapping Communities Reveal Rich Structure in Large-Scale Brain Networks during Rest and Task Conditions." *NeuroImage* 135 (July): 92–106.

Nanetti, Luca, Leonardo Cerliani, Valeria Gazzola, Remco Renken, and Christian Keysers. 2009. "Group Analyses of Connectivity-Based Cortical Parcellation Using Repeated K-Means Clustering." *NeuroImage* 47 (4): 1666–77.

Neubert, Franz-Xaver, Rogier B. Mars, Jérôme Sallet, and Matthew F. S. Rushworth. 2015. "Connectivity Reveals Relationship of Brain Areas for Reward-Guided Learning and Decision Making in Human and Monkey Frontal Cortex." *Proceedings of the National Academy of Sciences* 112 (20): E2695–2704.

Newman, M. E. J. 2006. "Modularity and Community Structure in Networks." *Proceedings of the National Academy of Sciences of the United States of America* 103 (23): 8577–82.

Nguyen, Vinh T., and Ross Cunnington. 2014. "The Superior Temporal Sulcus and the N170 during Face Processing: Single Trial Analysis of Concurrent EEG-fMRI." *NeuroImage* 86 (February): 492–502.

Nieto-Castanon, Alfonso, Satrajit S. Ghosh, Jason A. Tourville, and Frank H. Guenther. 2003. "Region of Interest Based Analysis of Functional Imaging Data." *NeuroImage* 19 (4): 1303–16.

Nishida, Mitsuhiro, Nikolaos Makris, David N. Kennedy, Mark Vangel, Bruce Fischl, Kalpathy S. Krishnamoorthy, Verne S. Caviness, and P. Ellen Grant. 2006. "Detailed Semiautomated MRI Based Morphometry of the Neonatal Brain: Preliminary Results." *NeuroImage* 32 (3): 1041–49.

Ogawa, S., T. M. Lee, A. R. Kay, and D. W. Tank. 1990. "Brain Magnetic Resonance Imaging with Contrast Dependent on Blood Oxygenation." *Proceedings of the National Academy of Sciences* 87 (24): 9868–72.

Ogawa, S., D. W. Tank, R. Menon, J. M. Ellermann, S. G. Kim, H. Merkle, and K. Ugurbil. 1992. "Intrinsic Signal Changes Accompanying Sensory Stimulation: Functional Brain Mapping with Magnetic Resonance Imaging." *Proceedings of the National Academy of Sciences* 89 (13): 5951–55.

Oishi, Kenichi, Karl Zilles, Katrin Amunts, Andreia Faria, Hangyi Jiang, Xin Li, Kazi Akhter, et al. 2008. "Human Brain White Matter Atlas: Identification and Assignment of Common Anatomical Structures in Superficial White Matter." *NeuroImage* 43 (3): 447–57.

Oliveira, Francisco P. M., and João Manuel R. S. Tavares. 2014. "Medical Image Registration: A Review." *Computer Methods in Biomechanics and Biomedical Engineering* 17 (2): 73–93.

Olszowy, Wiktor, John Aston, Catarina Rua, and Guy B. Williams. 2019. "Accurate Autocorrelation Modeling Substantially Improves fMRI Reliability." *Nature Communications* 10 (1): 1220.

Ono, Michio, Stefan Kubik, and Chad D. Abernathey. 1990. *Atlas of the Cerebral Sulci*. New York: Thieme Verlag.

Orban, Guy A., David Van Essen, and Wim Vanduffel. 2004. "Comparative Mapping of Higher Visual Areas in Monkeys and Humans." *Trends in Cognitive Sciences* 8 (7): 315–24.

Orban, Pierre, Julien Doyon, Michael Petrides, Maarten Mennes, Richard Hoge, and Pierre Bellec. 2015. "The Richness of Task-Evoked Hemodynamic Responses Defines a Pseudohierarchy of Functionally Meaningful Brain Networks." *Cerebral Cortex* 25 (9): 2658–69.

Parisot, Sarah, Salim Arslan, Jonathan Passerat-Palmbach, William M. Wells 3rd, and Daniel Rueckert. 2016. "Group-Wise Parcellation of the Cortex through Multi-Scale Spectral Clustering." *NeuroImage*, May. https://doi.org/10.1016/j.neuroimage.2016.05.035.

Parisot, Sarah, Ben Glocker, Markus D. Schirmer, and Daniel Rueckert. 2016. "GraMPa: Graph-Based
41